\documentclass[12pt, leqno]{article}
\usepackage{amsmath,setspace,multirow,lineno}
\usepackage[font=small,labelfont=bf]{caption}
\usepackage[top=1.2in, bottom=1.2in, left=1.18in, right=1.13in]{geometry}

\usepackage[round]{natbib}
\usepackage{color,soul}

\DeclareCaptionStyle{italic}[justification=centering]{labelfont={bf},textfont={it},labelsep=colon}
\usepackage{graphicx,psfrag,epsf,textcomp,epstopdf, amsthm}
\usepackage{enumerate, dsfont}
\usepackage{natbib}
\usepackage{url,xcolor}
\usepackage{booktabs, subfig, bm, paralist,mathpazo,tikz,todonotes,longtable,microtype,algorithm}

\usepackage[pdftex,colorlinks=true]{hyperref}
\definecolor{darkblue}{rgb}{0,0,.6}
\hypersetup{citecolor=darkblue,linkcolor=darkblue,urlcolor=darkblue}
\definecolor{DarkRed}{rgb}{.7,0,.4}

\usepackage{comment}

\newcommand{\blind}{0}

\newcommand{\X}{\mathcal{X}}
\newcommand{\Y}{\mathcal{Y}}

\graphicspath{{plots/}}

\newsavebox\CBox

 \newtheorem{@definition}{\sc Definition}[section]
 
 \newtheorem{proposition}{\sc Proposition}[section]

  \renewcommand\X{\mathcal{X}}

\begin{document}

\def\spacingset#1{\renewcommand{\baselinestretch}{#1}\small\normalsize} \spacingset{1}

\if0\blind
{
\title{\bf A partial least squares approach for function-on-function interaction regression}}
\author{
Ufuk Beyaztas\footnote{Postal address: Department of Economics and Finance, Pirireis University, Istanbul, Turkey; Email: ubeyaztas@pirireis.edu.tr, ORCID: 0000-0002-5208-4950}
\\
Department of Economics and Finance \\
Piri Reis University \\
\\
Han Lin Shang\\
Department of Actuarial Studies and Business Analytics \\
Macquarie University
}
\maketitle
\fi

\if1\blind
{
\title{\bf A partial least squares approach for function-on-function interaction regression}
} 
\maketitle
\fi

\begin{abstract}
A partial least squares regression is proposed for estimating the function-on-function regression model where a functional response and multiple functional predictors consist of random curves with quadratic and interaction effects. The direct estimation of a function-on-function regression model is usually an ill-posed problem. To overcome this difficulty, in practice, the functional data that belong to the infinite-dimensional space are generally projected into a finite-dimensional space of basis functions. The function-on-function regression model is converted to a multivariate regression model of the basis expansion coefficients. In the estimation phase of the proposed method, the functional variables are approximated by a finite-dimensional basis function expansion method. We show that the partial least squares regression constructed via a functional response, multiple functional predictors, and quadratic/interaction terms of the functional predictors is equivalent to the partial least squares regression constructed using basis expansions of functional variables. From the partial least squares regression of the basis expansions of functional variables, we provide an explicit formula for the partial least squares estimate of the coefficient function of the function-on-function regression model. Because the true forms of the models are generally unspecified, we propose a forward procedure for model selection. The finite sample performance of the proposed method is examined using several Monte Carlo experiments and two empirical data analyses, and the results were found to compare favorably with an existing method.
\end{abstract}


\newpage
\spacingset{1.56} 

\section{Introduction} \label{sec:intro}

Recent advances in computer storage and data collection have enabled researchers in diverse branches of science such as, for instance, chemometrics, meteorology, medicine and finance, recording data of characteristics varying over a continuum (time, space, depth, wavelength, etc). This is the case of data collected by spectrometer, rain gauges, electroencephalographs, or even just by means of high-performance computer. In all these cases, a number of subjects are observed densely over time, space or both. Through the application of interpolation or smoothing techniques, these data become functions. Thus, these data can be represented in a form of curve, image or shape. Functional data analysis has arisen as a field of statistics which provides statistical tools for analyzing this type of information \citep{ramsay2002, ramsay2006, ferraty2006, horvath2012, cuevas2014}. \cite{Hsing} and \cite{KoRe} provide a thorough overview of the research on theoretical results and case studies of functional data analysis methods. The focus of this study will be on constructing function-on-function regression models, where a functional response and multiple functional predictors both consist of random curves. 

Function-on-function regression models have been well studied in the literature \citep[see e.g.,][and references therein]{ramsay1991, yao2005, ramsay2006, MullerYao2008, matsui2009, he2010, wang2014, ivanescu2015, chiou2016}. The main objective of such models is to explain the association between the functional response and one or more functional predictors. Most of the function-on-function regression models that have been proposed concentrate on main effects only. Let $\Y(t)$ and $\X_m(s)$ ($m = 1, \cdots, M$) respectively denote a functional response and $M$ sets of functional predictors, where $s \in S$ and $t \in T$ are closed and bounded intervals on the real line. For ease of notation, $s$ and $t$ can be standardized to the interval $[0, 1]$. The function-on-function regression model with main effects is defined as follows:
\begin{equation}\label{eq:fof}
\Y(t) = \beta_0(t) + \sum_{m=1}^M \int_0^1 \X_m(s) \beta_m(s,t) ds + \epsilon(t),
\end{equation}
where $\beta_0(t)$ is the intercept function, $\beta_m(s,t)$ is the bivariate coefficient function, and $\epsilon(t)$ is the random error function with mean zero.

A regression model with only main effects is less flexible compared with the one that incorporates the quadratic term and interaction effects. A scalar-on-function regression model with the quadratic term of the functional predictor was proposed by \cite{YaoMuller2010}. This model was later extended to the case where there is more than one functional predictor by \cite{Fuchs} and \cite{Usset}, who consider interactions between multiple functional predictors. \cite{Matsui2019} proposed a simple function-on-function regression model with a quadratic term of the functional predictor as follows:
\begin{equation}\label{eq:Hdt}
\Y(t) = \beta_0(t) + \int_0^1 \X(s) \beta(s,t) ds + \int_0^1 \int_0^1 \X(s) \X(r) \gamma(s,r,t) ds dr  + \epsilon(t),
\end{equation}
where $\gamma(s,r,t)$ is the three-dimensional coefficient function of the quadratic term. \cite{Matsui2019} estimated model parameters by the penalized maximum likelihood (PML) method. \cite{SunWang} proposed a functional principal component (PC) regression to estimate the parameters in model~\eqref{eq:Hdt} and provided a procedure to test the significance of the quadratic term. \cite{LuoQi} proposed a multiple function-on-function regression model with quadratic and interaction effects as follows:
\begin{equation}\label{eq:main}
\Y(t) = \beta_0(t) + \sum_{m \in \mathcal{M}} \int_0^1 \X_m(s) \beta_m(s,t) ds + \sum_{m,n \in \mathcal{I}} \int_0^1 \int_0^1 \X_m(s) \X_n(r) \gamma_{mn}(s,r,t) ds dr + \epsilon(t),
\end{equation}
where the error function $\epsilon(t)$ has mean zero and is independent of $\X_m(s)$'s, and $\mathcal{M}$ and $\mathcal{I}$, respectively, denote the index sets of main and quadratic/interaction effects. The index sets $\mathcal{M}$ and $\mathcal{I}$ are assumed to be any subset of $\left\lbrace 1, \cdots, M \right\rbrace$ and $\left\lbrace \left( m,n \right): 1 \leq m \leq n \leq M \right\rbrace$, respectively. In addition, \cite{LuoQi} proposed a stepwise procedure for model selection. The numerical results produced from the previous studies have shown that functional regression models including quadratic term and interaction effects perform better than standard functional regression models in the presence of interaction.

We propose a functional partial least squares (PLS) regression to estimate the multiple function-on-function regression model with quadratic and interaction effects given in~\eqref{eq:main}. Several functional PLS methods have been proposed to estimate parameters of the functional regression models. For example, \cite{PredSap}, \cite{ResiOgden}, \cite{Kramer}, \cite{Agu2010}, and \cite{Agu2016} proposed functional PLS methods for scalar-on-function regression models. \cite{Delaigle2012a} examined the theoretical properties of functional PLS methods in scalar-on-function regression models. In addition, \cite{Febrero} compared the finite sample performances of these methods and discussed their advantages and disadvantages. \cite{HydSh2009} proposed a functional PLS regression based on a lagged functional predictor and a functional response for forecasting functional time series. \cite{Escabias2007} proposed a functional PLS method for logit regression. Functional PLS has also been used for other regression problems. For example, \cite{ferraty2006} defined a functional PLS-based semi-metric for nonparametric functional predictor variables and \cite{Preda2007} and \cite{Delaigle2012b} proposed functional PLS-based functional data classification methods. Our method differs from these existing functional PLS methods in terms of the number of functional predictors in the model: while the existing functional PLS methods allow only one functional predictor in the model, the proposed method allows for more than one functional predictor and their quadratic/interaction effects. The main contributions of the proposed method include:
\begin{inparaenum}
\item[i)] vectorization of function values,
\item[ii)] an application of PLS (for vector-valued predictors and responses), and 
\item[iii)] a forward stepwise variable selection.
\end{inparaenum}
The interaction terms, including the quadratic terms, are treated as linear terms once they are appropriately transformed.

We use the idea of functional PLS proposed by \cite{PredSap} and \cite{Agu2010} to estimate model~\eqref{eq:main} in Section~\ref{sec:methodology}. In our proposed method, all the functional predictors and their quadratic/interaction terms are first stacked into a two-dimensional function ($\pmb{\X}(s,r)$; see~\eqref{main_ed}). Then, a functional PLS regression of $\Y(t)$ on $\pmb{\X}(s,r)$ is considered. Functional random variables, by their nature, belong to infinite-dimensional space. Thus, the direct estimation of function-on-function regression models is generally an ill-posed problem. One remedy to estimate function-on-function regression models is the approximation of functional data via finite-dimensional basis function expansion methods. In this way, a function-on-function regression model is converted into a multivariate regression model of basis expansion coefficients. Then, the functional PLS regression of $\Y(t)$ on $\pmb{\X}(s,r)$ is estimated using the basis expansion approximations of $\Y(t)$ and $\pmb{\X}(s,r)$. We show the equality of the functional PLS and the PLS regression constructed via basis expansions of $\Y(t)$ and $\pmb{\X}(s,r)$ using the metrics associated with basis functions. In addition, we provide an explicit formula for the coefficient function of the functional PLS regression model from the PLS estimate of the regression model constructed via basis coefficient expansions. When considering model~\eqref{eq:main}, the form of the true model is usually unspecified and the index sets $\mathcal{M}$ and $\mathcal{I}$ are unknown in practice \citep{LuoQi}. To this end, we further propose a forward procedure to determine the significant functional predictors and their quadratic/interaction terms. Our numerical results, which will be discussed in detail in Sections~\ref{sec:results} and~\ref{sec:real}, reveal that the proposed method based on the forward procedure produces better prediction performance compared with the full model, which includes all the predictors and their quadratic and interaction terms. Section~\ref{sec:conc} concludes the paper.

\section{Methodology} \label{sec:methodology}

Let us consider a set of random functional response $\{Y_1(t), \dots, Y_N(t)\}$ containing $N$ independent realization $\Y_i$ and $M$ sets of random functional predictors $\{\X_{mi}(s); m = 1, \cdots, M, i=1, \cdots, N\}$. With these realizations, we aim to construct the multiple function-on-function regression model with quadratic and interaction effects given in~\eqref{eq:main}. We assume that the functional random variables are the elements of square-integrable and real-valued functions $\mathcal{L}_2[0,1]$, and are of second-order stochastic processes. Without loss of generality, we assume that $\mathbb{E} \left[ \Y(t) \right] = 0$ and $\mathbb{E} \left[ \X_m(s) \right] = 0$ for $m = 1, \cdots, M$ and $\forall t,s \in [0,1]$, and the functional data pairs $\left\lbrace \X_{mi}(s), Y_i(t) \right\rbrace$ are fully observed. For the sake of clarity, we consider the full model where all the functional predictors and their quadratic/interaction terms are included in the model; that is $m = 1, \cdots, M$ and $(m,n) = (1,1), (1,2), \cdots, (1,M),\allowbreak (2,3), \cdots, (M-1, M), (M,M)$. Using the continuity of functional objects, we express model~\eqref{eq:main} as follows:
\begin{align}
\Y(t) &= \sum_{m \in \mathcal{M}} \int_0^1 \X_m(s) \beta_m(s,t) ds + \sum_{m \in \mathcal{M}} \sum_{n \in \mathcal{M}} \int_0^1 \int_0^1 \X_m(s) \X_n(r) \gamma_{mn}(s,r,t) ds dr + \epsilon(t), \nonumber \\
\Y(t) &= \int_0^1 \sum_{m \in \mathcal{M}} \X_m(s) \beta_m(s,t) ds + \int_0^1 \sum_{m \in \mathcal{M}} \X_m(s) \int_0^1 \sum_{n \in \mathcal{M}} \X_n(r) \gamma_{mn}(s,r,t) ds dr + \epsilon(t), \nonumber \\
\Y(t) &= \int_0^1 \sum_{m \in \mathcal{M}} \X_m(s) \left[ \beta_m(s,t) + \int_0^1 \sum_{n \in \mathcal{M}} \X_n(r) \gamma_{mn}(s,r,t) dr \right] ds + \epsilon(t), \nonumber \\
\Y(t) &= \sum_{m \in \mathcal{M}} \int_0^1 \X_m(s) \left[ \beta_m(s,t) +  \sum_{n \in \mathcal{M}} \int_0^1 \X_n(r) \gamma_{mn}(s,r,t) dr \right] ds + \epsilon(t). \nonumber
\end{align}

Let $\pmb{\X}(s,r): \mathbb{R}^2 \rightarrow \mathbb{R}^{M (M+3) / 2}$ be a two-dimensional vector-valued function as follows:
\begin{align*}
\pmb{\X}(s,r) &= (\X_1(s), \cdots, \X_M(s) \\
& \X_1(s) \X_1(r), \X_1(s) \X_2(r), \cdots, \X_1(s) \X_M(r) \\
& X_2(s) \X_2(r), \cdots, X_{M-1}(s) X_M(r), \X_M(s) \X_M(r) )^\top
\end{align*}
Herein, the $\pmb{\X}(s,r)$ is considered as a single two-dimensional vector-valued function but including $M (M+3)/2$ ($M$ sets of functional predictors and $M \times (M+1)/2$ sets of functional quadratic/interaction terms) sets of functional predictors and their quadratic/interaction terms. Hereafter, we will consider the following regression model:
\begin{equation}\label{main_ed}
\Y(t) = \int_0^1 \int_0^1 \pmb{\X}(s,r) \Xi(s,r,t) ds dr + \epsilon(t),
\end{equation}
where $\Xi(s,r,t)$ is a single three-dimensional coefficient function comprising $M (M+3)/2$ sets of coefficient functions of both main and quadratic/interaction effects, as follows:
\begin{align*}
\Xi(s,r,t) &= (\beta_1(s,t), \cdots, \beta_M(s,t) \\
& \gamma_{11}(s,r,t), \gamma_{12}(s,r,t), \cdots, \gamma_{1M}(s,r,t) \\
& \gamma_{22}(s,r,t), \cdots, \gamma_{M-1,M}(s,r,t), \gamma_{MM}(s,r,t))^\top
\end{align*}
The coefficient identifiability of~\eqref{main_ed} is discussed in Appendix.

In practice, the random functions of a functional dataset $\left\lbrace x_i(t): i = 1, \cdots, N \right\rbrace$ are observed in the set of finite discrete time points $\left\lbrace t_{\ell}: \ell = 1, \cdots, L \right\rbrace$; that is $\left\lbrace x_i(t_{\ell}) \right\rbrace$. Thus, the discretely observed data points are first converted into their functional forms using a suitable method such as basis function expansion \citep{ramsay2006} and nonparametric smoothing of functions \citep{ferraty2006} before fitting the functional regression model. In this study, we consider the basis function expansion approach. Briefly, a random function $x(t)$ can be approximated by the linear combinations of basis functions, $\phi_k(t)$, and associated coefficients, $c_{k}$, for a sufficiently large number of basis functions, $K$:
\begin{equation*}
x(t) \approx \sum_{k=1}^K c_k \phi_k(t).
\end{equation*}
Several basis function expansion methods such as $B$-spline, Fourier, radial, and wavelet bases have been proposed \citep[see, e.g.,][]{ramsay2006}. In this study, we consider the $B$-spline basis function expansion method. 

Let $K_{\Y}$ and $K_{\pmb{\X}}$ denote the number of basis functions used for approximating $\Y(t)$ and $\pmb{\X}(s,r)$, respectively. They are expressed by basis function expansion as follows:
\begin{align}
\Y(t) &= \sum_{k=1}^{K_{\Y}} c_k \phi_k(t) = \pmb{c}^\top \pmb{\Phi}(t), \label{bfe:y} \\
\pmb{\X}(s,r) &= \sum_{j=1}^{K_{\pmb{\X}}} \sum_{l=0}^{K_{\pmb{\X}}} d_{jl} \psi_{jl}(s,r) = \pmb{d}^\top \pmb{\Psi}(s,r), \label{bfe:x}
\end{align} 
where $\pmb{\Phi}(t)$ and $\pmb{\Psi}(s,r)$ respectively are the $K_{\Y}$ and $K_{\pmb{\X}} \times M (M+3)/2$ dimensional basis functions vector/matrix, as follows:
\begin{align*}
\pmb{\Phi}(t) &= (\phi_1(t), \cdots, \phi_{K_{\Y}}(t))^\top \\
\pmb{\Psi}(s,r) &= (\psi_1(s), \cdots, \psi_M(s)\\
& \psi_1(s) \psi_1(r), \psi_1(s) \psi_2(r), \cdots, \psi_1(s), \psi_M(r) \\
& \psi_2(s) \psi_2(r), \cdots, \psi_{M-1}(s) \psi_M(r), \psi_M(s) \psi_M(r))^\top
\end{align*}
with $\psi_i(s) = \left( \psi_{i1}, \cdots, \psi_{iK_{\pmb{\X}}} \right)^\top$ and $\pmb{c}$ and $\pmb{d}$ are the $K_{\Y}$ and $K_{\pmb{\X}}$ dimensional coefficient vectors, respectively. In addition, let us assume that the three-dimensional coefficient function $\Xi(s,r,t)$ in~\eqref{main_ed} is expressed by basis expansion as follows:
\begin{equation}\label{bfe:xi}
\Xi(s,r,t) = \sum_{j=1}^{K_{\pmb{\X}}} \sum_{l=1}^{K_{\pmb{\X}}} \sum_{k=1}^{K_{\Y}} \xi_{jlk} \psi_{jl}(s,r) \phi_k(t) = \pmb{\Psi}^\top(s,r) \Gamma_{(3)}^\top \pmb{\Phi}(t),
\end{equation}
where $\pmb{\Gamma}_{(3)}$ is a $K_{\Y} \times K_{\pmb{\X}}^2$ dimensional matrix obtained by matricizing a three-dimensional $K_{\pmb{\X}} \times K_{\pmb{\X}} \times K_{\Y}$ tensor $\underline{\pmb{\Gamma}} = \left( \xi_{jlk} \right)_{jlk}$ with respect to the 3\textsuperscript{rd} array \citep{Lathuawer, Matsui2019}. Without loss of generality, we further assume that the error function in~\eqref{main_ed} is represented as a basis function expansion as follows:
\begin{equation*}
\epsilon(t) = \pmb{e} \pmb{\Phi}(t),
\end{equation*}
where $\pmb{e} = \left( \pmb{e}_1, \cdots, \pmb{e}_N \right)^\top$ is an $N$-dimensional vector comprising random variables $\pmb{e}_i = \left(e_{i1}, \cdots, e_{iK_{\Y}} \right)^\top$. Now, replacing~\eqref{bfe:y} to~\eqref{bfe:xi} with~\eqref{main_ed} yields:
\begin{align}
\pmb{c}^\top \pmb{\Phi}(t) &= \pmb{d}^\top \pmb{\Psi}(s,r) \pmb{\Psi}^\top(s,r) \Gamma_{(3)}^\top \pmb{\Phi}(t) + \pmb{e} \pmb{\Phi}(t), \nonumber \\
\pmb{c} &= \pmb{d}^\top \pmb{\Psi} \Gamma_{(3)}^\top + \pmb{e}, \label{discrete}
\end{align}
where $\pmb{\Psi} = \int_0^1 \int_0^1 \pmb{\Psi}(s,r) \pmb{\Psi}^\top(s,r) ds dr$. Thus, the problem of estimating the function-on-function regression model in~\eqref{main_ed} is reduced to estimating the multivariate regression model~\eqref{discrete} with coefficient matrix $\Gamma_{(3)}^\top$. 

When considering~\eqref{discrete}, the dimensions of the matrices increase exponentially when a large number of predictors are used in the model or when a large number of basis functions are used to approximate the functional objects. In such cases, an ill-posed problem (multicollinearity) arises and estimating $\Gamma_{(3)}^\top$ using traditional methods, such as least squares and maximum likelihood results in infinitely many solutions. Regularisation methods such as the PML method proposed by \cite{Matsui2019} may be used to overcome this problem and to produce an accurate estimate for $\Gamma_{(3)}^\top$. However, it is computationally time-consuming, and it may not be possible to obtain the PML estimate of $\Gamma_{(3)}^\top$ using a computer with standard memory. One remedy to obtain a stable estimate for $\Gamma_{(3)}^\top$ is to use dimension-reduction techniques, such as PC or PLS regression models. PC regression has been proposed by \cite{AguileraPCA} and \cite{yao2005} to estimate standard function-on-function regression models and by \cite{SunWang} to estimate univariate function-on-function quadratic regression models. However, this method does not take into account the relationship between response and predictor variables when deciding the PCs. Further \cite{Agu2010} showed that in the functional case, the parameter function estimated with PLS is much more accurate than with PC. Therefore, we consider functional PLS regression to estimate model~\eqref{main_ed}.

\subsection{PLS for the functional interaction regression model}

In the multivariate case, the main goal of PLS regression is to construct a linear model between an $n \times m$ dimensional fixed full-rank design matrix $\pmb{X} = \left( X_1, \cdots, X_m\right)^\top$ and an $n \times r$ dimensional matrix for response variables $\pmb{Y} = \left( Y_1, \cdots, Y_r \right)^\top$ as follows:
\begin{equation*}
\pmb{Y} = \pmb{X} \pmb{\beta} + \pmb{E},
\end{equation*}
where $\pmb{\beta}$ and $\pmb{E}$ are the $m \times r$ dimensional coefficient matrix and the error matrix with dimensions $n \times r$, respectively. The PLS regression produces orthogonal latent components $\pmb{\mathbb{T}}$ formed as linear combinations of predictor matrix $\pmb{X}$ and an appropriate weight matrix $\pmb{W}$ such that $\pmb{\mathbb{T}} = \pmb{X} \pmb{W}$. Herein, the weight matrix is computed so that it maximizes the covariance between $\pmb{Y}$ and $\pmb{\mathbb{T}}$. Then, an ordinary least squares regression is used to capture the relationship between $\pmb{Y}$ and $\pmb{\mathbb{T}}$ such that $\pmb{Y} = \pmb{\mathbb{T}} \pmb{Q} + \pmb{E}$. In turn, it produces the loading matrix $\pmb{Q}$ for $\pmb{Y}$. Then, the PLS regression is constructed as follows:
\begin{equation*}
\pmb{Y} = \pmb{X} \pmb{\Omega} + \pmb{E},
\end{equation*}
where $\pmb{\Omega} = \pmb{W} \pmb{Q}$.

Several functional versions of PLS regression have been proposed to explore the relationship between a functional predictor and a scalar and/or functional response \citep[see e.g.,][]{PredSap, HydSh2009, Agu2010, Agu2016, BS19}. Similar to the discrete case, the functional PLS method produces orthogonal latent components as linear combinations of the functional predictors by maximizing the squared covariance between the response and orthogonal latent components \citep{Tucker}. The PLS components are determined by an iterative process. In this study, we extend the functional PLS methods proposed by \cite{PredSap} and \cite{Agu2010} to estimate the model given in~\eqref{main_ed}. Our extension differs from the methods of \cite{PredSap} and \cite{Agu2010} in the context of the number of functional predictors in the model. The original functional PLS methods allow for only one functional predictor in the model. However, our proposed method is possible to construct a regression model with more than one functional predictor and their quadratic/interaction effects. The proposed method first stacks all the main and quadratic/interaction effects into a two-dimensional function, $\pmb{\X}(s,r)$, it then applies a PLS method on the vectorized functional variables to estimate the model parameters.

The PLS components of regression model~\eqref{main_ed} can be obtained as solutions of Tucker's criterion extended to functional-type data as follows:
\begin{equation*}
\underset{\begin{subarray}{c}
  \kappa \in \mathcal{L}_2([0,1] \times [0,1]),~ \Vert \kappa \Vert_{\mathcal{L}_2[0,1] \times [0,1]} = 1 \\
  \zeta \in \mathcal{L}_2[0,1],~ \Vert \zeta \Vert_{\mathcal{L}_2[0,1]} = 1
  \end{subarray}}{\max} \text{Cov}^2 \left( \int_0^1 \int_0^1 \pmb{\X}(s,r) \kappa(s,r) ds dr, ~ \int_0^1 \Y(t) \zeta(t) dt \right),
\end{equation*}
where $\left\Vert \cdot \right\Vert_{\mathcal{L}_2}$ denotes $\mathcal{L}_2$ norm, which is approximated by the Riemann sum \citep{LuoQi}. The functional PLS components are also the eigenvectors of the Escoufier operator \citep{Escoufier}. Let $W^{\pmb{\X}}$ and $W^{\Y}$, respectively, denote the Escoufier's operators associated with $\pmb{\X}(s,r)$, with respect to $\Y(t)$ as follows:
\begin{equation*}
W^{\pmb{\X}} = \int_0^1 \int_0^1 \mathbb{E} \left[ \pmb{\X}(s,r) Z \right] \pmb{\X}(s,r) ds dr, \qquad W^{\Y} = \int_0^1 \mathbb{E} \left[ \Y(t) Z \right] \Y(t) dt, \qquad \forall Z \in \mathcal{L}_2[0,1].
\end{equation*}
Then, the first PLS component, say $\eta_1$, of model~\eqref{main_ed} is equal to the eigenvector corresponding to the largest eigenvalue ($\lambda_{\max}$) of $W^{\pmb{\X}} W^{\Y}$:
\begin{equation*}
W^{\pmb{\X}} W^{\Y} \eta_1 = \lambda_{\max} \eta_1.
\end{equation*}
Accordingly, the first PLS component is defined as follows:
\begin{equation*}
\eta_1 = \int_0^1 \int_0^1 \kappa_1(s,r) \pmb{\X}(s,r) ds dr,
\end{equation*}
where the weight function $\kappa_1(s,r)$ associated with $\eta_1$ is given by:
\begin{equation*}
\kappa_1(s,r) = \frac{\int_0^1 \mathbb{E} \left[ \Y(t) \pmb{\X}(s,r)\right] dt}{\sqrt{\int_0^1 \int_0^1 \left( \int_0^1 \mathbb{E} \left[ \Y(t) \pmb{\X}(s,r)\right] dt \right)^2 ds dr}}.
\end{equation*}

PLS is an iterative stepwise procedure. Let $h = 1, 2, \cdots,$ denote the iteration number. Define $\pmb{\X}_h(s,r)$ and $\Y_h(t)$ by the residuals obtained from the following regressions:
\begin{align*}
\pmb{\X}_h(s,r) &= \pmb{\X}_{h-1}(s,r) - p_h(s,r) \eta_h, \\
\Y_h(t) &= \Y_{h-1}(t) - \zeta_h(t) \eta_h,
\end{align*}
where $\pmb{\X}_0(s,r) = \pmb{\X}(s,r)$, $\Y_0(t) = \Y(t)$, $p_h(s,r) = \frac{\mathbb{E} \left[ \pmb{\X}_{h-1}(s,r) \eta_h\right]}{\mathbb{E} \left[ \eta_h^2 \right]}$, and $\zeta_h(t) = \frac{\mathbb{E} \left[ \Y_{h-1}(t) \eta_h \right]}{\mathbb{E} \left[ \eta_h^2 \right]}$. Then, the $h^{\text{th}}$ PLS component, $\eta_h$, is equal to the eigenvector corresponding to largest eigenvalue of $W_{h-1}^{\pmb{\X}} W_{h-1}^{\Y}$:
\begin{equation*}
W_{h-1}^{\pmb{\X}} W_{h-1}^{\Y} \eta_h = \lambda_{\max} \eta_h,
\end{equation*}
where $W_{h-1}^{\pmb{\X}}$ and $W_{h-1}^{\Y}$ are the Escoufier's operators associated with $\pmb{\X}_{h-1}(s,r)$ and $\Y_{h-1}(t)$, respectively. That is, $\eta_h$ is given as follows:
\begin{equation*}
\eta_h = \int_0^1 \int_0^1 \kappa_h(s,r) \pmb{\X}_{h-1}(s,r) ds dr,
\end{equation*}
where the weight function associated with $\eta_h$ is given by:
\begin{equation*}
\kappa_h(s,r) = \frac{\int_0^1 \mathbb{E} \left[ \Y_{h-1}(t) \pmb{\X}_{h-1}(s,r)\right] dt}{\sqrt{\int_0^1 \int_0^1 \left( \int_0^1 \mathbb{E} \left[ \Y_{h-1}(t) \pmb{\X}_{h-1}(s,r)\right] dt \right)^2 ds dr}}.
\end{equation*}

Following \cite{Agu2010}, the properties of the PLS components of model~\eqref{main_ed} can be summarized by the following proposition:
\begin{proposition}\label{prp:1}
For any $h \geq 1$
\begin{itemize}
\item[i)] $\left\lbrace \eta_h \right\rbrace_{h \geq 1}$ forms an orthogonal system in the linear space spanned by $\pmb{\X}(s,r)$,
\item[ii)] $\Y(t) = \zeta_1(t) \eta_1 + \cdots + \zeta_h(t) \eta_h + \Y_h(t)$,
\item[iii)] $\pmb{\X}(s,r) = p_1(s,r) \eta_1 + \cdots + p_h(s,r) \eta_h + \pmb{\X}_h(s,r)$,
\item[iv)] $\mathbb{E} \left[ \Y_h(t) \eta_j \right] = 0$, $\forall j = 1, \cdots, h$,
\item[v)] $\mathbb{E} \left[ \pmb{\X}_h(s,r) \eta_j \right] = 0$, $\forall j = 1, \cdots, h$.
\end{itemize}
\end{proposition}

\subsection{PLS for basis expansion of the functional interaction regression model}

Functional PLS components are estimated using an observed sample of the functional dataset. However, the random functions of functional variables are observed in the set of finite discrete time points. In this case, the main problem related to the sample estimation of PLS components is that of estimating the Escoufier's operators from discrete time observations \citep{Agu2010}. To overcome this problem, we consider the basis expansions of the observed functional dataset, and construct the PLS regression using the basis expansion coefficients of $\Y(t)$ and $\pmb{\X}(s,r)$. Let us revisit model~\eqref{main_ed} and its basis function expansion in~\eqref{discrete}. Then, we provide the following proposition to show the relationship between the functional PLS of $\Y(t)$ on $\pmb{\X}(s,r)$ and the PLS of $\pmb{c}$ on $\pmb{d}$. The proof of the proposition has been relegated to the Appendix.

\begin{proposition}\label{prop:2}
Under the assumption that the functional variables $\Y(t)$ and $\pmb{\X}(s,r)$ are spanned with finite number of basis functions, let $\pmb{\Phi}$ and $\pmb{\Psi}$, respectively, denote $K_{\Y} \times K_{\Y}$ and $K_{\pmb{\X}}^2 \times K_{\pmb{\X}}^2$ dimensional symmetric matrices with entries the inner products of the basis functions $\int_0^1 \pmb{\Phi}(t) \pmb{\Phi}^\top(t) dt$ and $\int_0^1 \int_0^1 \pmb{\Psi}(s,r) \pmb{\Psi}^\top(s,r) ds dr$, respectively. Also, let $\pmb{\Phi}^{1/2}$ and $\pmb{\Psi}^{1/2}$ denote the square roots of $\pmb{\Phi}$ and $\pmb{\Psi}$, respectively. Then, the functional PLS regression of $\Y(t)$ on $\pmb{\X}(s,r)$ is equivalent to the PLS regression of $\pmb{\Phi}^{1/2} \pmb{c}$ on  $\pmb{\Psi}^{1/2} \pmb{d}$ in the sense that at each step $h$ of the PLS algorithm, the same PLS components are obtained for both PLS regressions. 
\end{proposition}

The PLS approximation of the coefficient function $\Xi(s,r,t)$ in~\eqref{main_ed} can be expressed in the bases $\pmb{\Phi}(t)$ and $\pmb{\Psi}(s,r)$. Denote by $\pmb{\Theta}^h = \left( \Theta_1^h, \cdots, \Theta_{K_{\pmb{\X}}^2}^h \right)$ the regression coefficients of $\pmb{\Phi}^{1/2} \pmb{c}$ on $\pmb{\Psi}^{1/2} \pmb{d}$ obtained by the PLS regression at step $h$. Then, we have
\begin{align*}
\pmb{\Phi}^{1/2} \pmb{c} &= \pmb{d} \pmb{\Psi}^{1/2} \pmb{\Theta}^h, \\
\widehat{\Y}(t) &= \int_0^1 \int_0^1 \pmb{\X}(s,r) \widehat{\Xi}(s,r,t) ds dr,
\end{align*}
where 
\begin{equation*}
\widehat{\Xi}(s,r,t) = \sum_{j=1}^{K_{\pmb{\X}}} \sum_{l=1}^{K_{\pmb{\X}}} \sum_{k=1}^{K_{\Y}} \left( \left( \pmb{\Psi}^{1/2} \right)^{-1} \pmb{\Theta}^h \left( \pmb{\Phi}^{1/2} \right)^{-1} \right)_{(jl)} \psi_{jl}(s,r) \phi_k(t), \qquad s,r,t \in \mathcal{L}_2[0,1].
\end{equation*}
Herein, $\widehat{\Xi}(s,r,t)$ denotes the PLS approximation of the three-dimensional coefficient function $\Xi(s,r,t)$. These results demonstrate that the infinite-dimensional estimation problem can be reduced to a simple finite-dimensional PLS regression setting using the particular metrics $\pmb{\Phi}$ and $\pmb{\Psi}$ in the spaces of expansion coefficients $\pmb{c}$ and $\pmb{d}$, respectively. Through the basis expansion, one can implement a finite multivariate PLS regression setting and a PLS algorithm, such as nonlinear iterative partial least squares (NIPALS) of \cite{nipals}, the SIMPLS of de \cite{simpls}, and the improved kernel PLS of \cite{dayal} to approximate the three-dimensional coefficient function $\Xi(s,r,t)$. In this study, the NIPALS algorithm is used to obtain the PLS approximation of $\Xi(s,r,t)$.

\subsection{Model selection}

\subsubsection{Variable selection procedure}\label{sec:vsp}

The functional regression model in~\eqref{main_ed} includes too many interaction terms when a large number of functional predictors are used in the model. However, not all the functional terms may be significant for the model. Thus, a variable selection procedure is needed to select the best subset of functional predictors and quadratic/interaction terms. In this context, several criteria, such as the Akaike information criterion, Bayesian information criterion, and the predicted residual error sum of squares, have been proposed. However, a variable selection procedure based on these criteria will be computationally intensive, since the number of all possible models is very large when there are a large number of predictors in the model. Various variable selection procedures, such as backward elimination and forward selection, have been proposed to overcome this problem. In this study, we propose the following forward selection procedure to select the significant functional predictors and quadratic/interaction terms.
\begin{itemize}
\item[Step 1.] We consider selecting the main effect terms first, since they are generally more important for the model compared with the quadratic term and interaction effects. To start with, first, $M$ function-on-function regression models, each of which includes the common response and a predictor, are constructed as follows:
\begin{equation*}
\Y(t) = \int_0^1 \int_0^1 \pmb{\X}_m(s,r) \Xi_m(s,r,t) ds dr + \epsilon_m(t),
\end{equation*}
where $\pmb{\X}_m(s,r) = \X_m(s)$ and $r = 0$ for $m = 1, \cdots, M$. Among these models, the one having the smallest mean squared error (MSE),
\begin{equation*}
\text{MSE} = \frac{1}{N}\sum_{j=1}^{N} \left\Vert \Y_j(t) - \widehat{\Y}_j(t) \right\Vert^2_{\mathcal{L}_2},
\end{equation*}
where $\widehat{\Y}_j(t)$ is the prediction of $\Y_j(t)$ for $j = 1, \cdots, N$, is chosen as an initial model. Let $\pmb{\X}^{(1)}(s,r)$ and $\text{MSE}^{(1)}$ denote the predictor variable in the initial model and the MSE obtained from this model, respectively. Then, $M-1$ function-on-function regression models, each of which includes the response and a predictor, are constructed as follows:
\begin{equation*}
\Y(t) = \int_0^1 \int_0^1 \pmb{\X}_m(s,r) \Xi_m(s,r,t) ds dr + \epsilon_m(t),
\end{equation*}
where $\pmb{\X}_m(s,r) = \left\lbrace \pmb{\X}^{(1)}(s,r), \X_m(s) \right\rbrace$, $\X_m(s) \neq \pmb{\X}^{(1)}(s,r)$, and $r = 0$ for $m = 1, \cdots, M-1$, and the MSE is calculated for each of these models. The predictor variable that produces the smallest MSE, $\pmb{\X}^{(2)}(s,r)$, is chosen as the predictor variable for the current model if $\text{MSE}^{(2)} < \text{MSE}^{(1)}$ where $\text{MSE}^{(2)}$ is the calculated MSE when $\pmb{\X}^{(2)}(s,r)$ is used to estimate the current model. This process is repeated until all the significant variables are included in the model.
\item[Step 2.] We consider an iterative procedure similar to that described in Step 1 to select the significant quadratic and interaction effects terms. Let $\mathcal{M}^*$ ($\mathcal{M}^* \subseteq \mathcal{M}$) denote the index set of all significant main effect terms found in the previous step. Denote by $M^*$ the number of items in $\mathcal{M}^*$, so there are $M^* \times (M^* + 1) / 2$ quadratic/interaction terms to be considered. Let $\text{MSE}^{(\mathcal{M}^*)}$ denote the calculated MSE when all the significant main effect terms are included in the model. To select quadratic/interaction effect terms, first, $M^* \times (M^* + 1) / 2$ function-on-function regression models including the response and a predictor are constructed as follows:
\begin{equation*}
\Y(t) = \int_0^1 \int_0^1 \pmb{\X}_m(s,r) \Xi_m(s,r,t) ds dr + \epsilon_m(t),
\end{equation*}
where $\pmb{\X}_m(s,r) = \left[ \left\lbrace \X_m(s) \right\rbrace_{m \in \mathcal{M}^*} \left(1 + \X_n(r) \right) \right]$ for $n = 1, \cdots, M^*$, and the MSE is calculated for each of these models. Then, the first quadratic/interaction effect term that leads to the greatest improvement to the $\text{MSE}^{(\mathcal{M}^*)}$ is included in the model. Other quadratic/interaction effect terms are iteratively selected to enter the model based on their improvements to the MSE of the current model.
\end{itemize}

\subsubsection{Selection of the optimum number of PLS components}

To select the optimum number of PLS components, we consider a cross-validation-based mean squared prediction error (MSPE) as follows:
\begin{itemize}
\item[1)] First, the data are randomly divided into training and testing samples with sample sizes $N_{train}$ and $N_{test}$, respectively. Note that in our numerical analyses, the training and testing samples are determined based on the 50\% -- 50\% rule. 
\item[2)] Using the training set, $B_h$ PLS regression models based on $h = 1, \cdots, B_h$ number of PLS components are constructed.
\item[3)] The values of the response variable in the testing sample are predicted using the values of the predictor variables in the testing sample and the coefficient matrices obtained from the previous step, and the MSPE is computed for each $B_h$:
\begin{equation*}
\text{MSPE}(h) = \frac{1}{N_{test}} \sum_{i=1}^{N_{test}} \left\Vert \Y_i(t) - \widehat{\Y}_i^h(t) \right\Vert^2_{\mathcal{L}_2},
\end{equation*}
where $\widehat{\Y}_i^h(t)$ is the prediction of $\Y_i(t)$ based on $h$ PLS components.
\end{itemize}
Based on the above algorithm, the optimum number of PLS components is equivalent to one which produces the minimum MSPE. Note that the cross-validation approach should only be applied to the training set to select the number of components. Within the training set, we can further split the samples into a training and a test sample. We note that our procedure for the selection of the optimum number of PLS components consists of two steps. In the first step, the significant main, quadratic, and interaction effects using a fixed number of PLS components, say 8, are determined as discussed in Section~\ref{sec:vsp}. Then, in the second step, the cross-validation-based MSPE criterion given above is used to determine the number of optimum PLS components. Our numerical analyses (not reported in the paper) have shown that the choice of a fixed number of PLS components in the first step does not have a significant effect on the determination of main, quadratic, and interaction effects when $h \geq 4$.

\section{Numerical results} \label{sec:results}

\begin{figure}[!htbp]
  \centering
  \includegraphics[width=5.9cm]{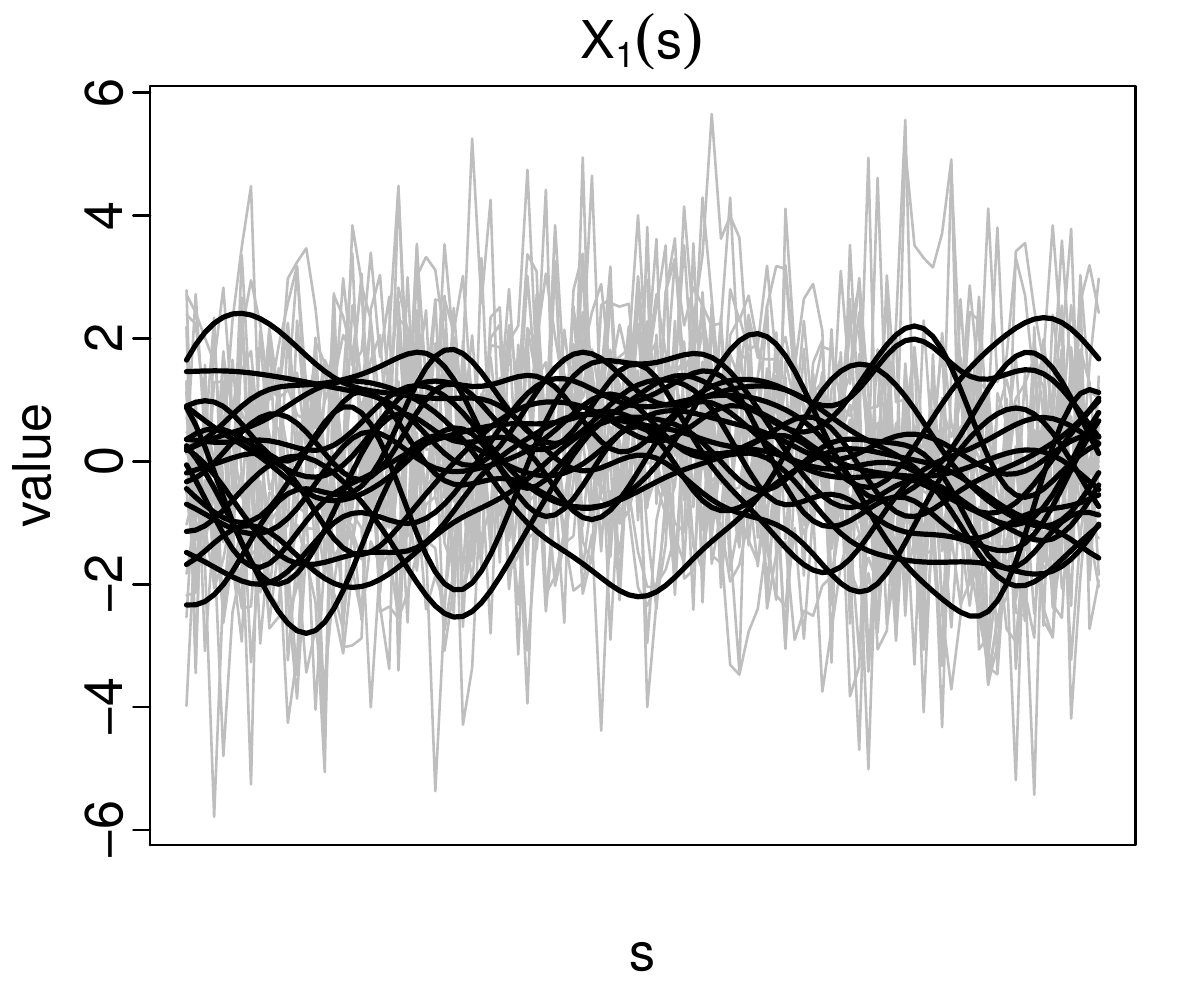}
  \includegraphics[width=5.9cm]{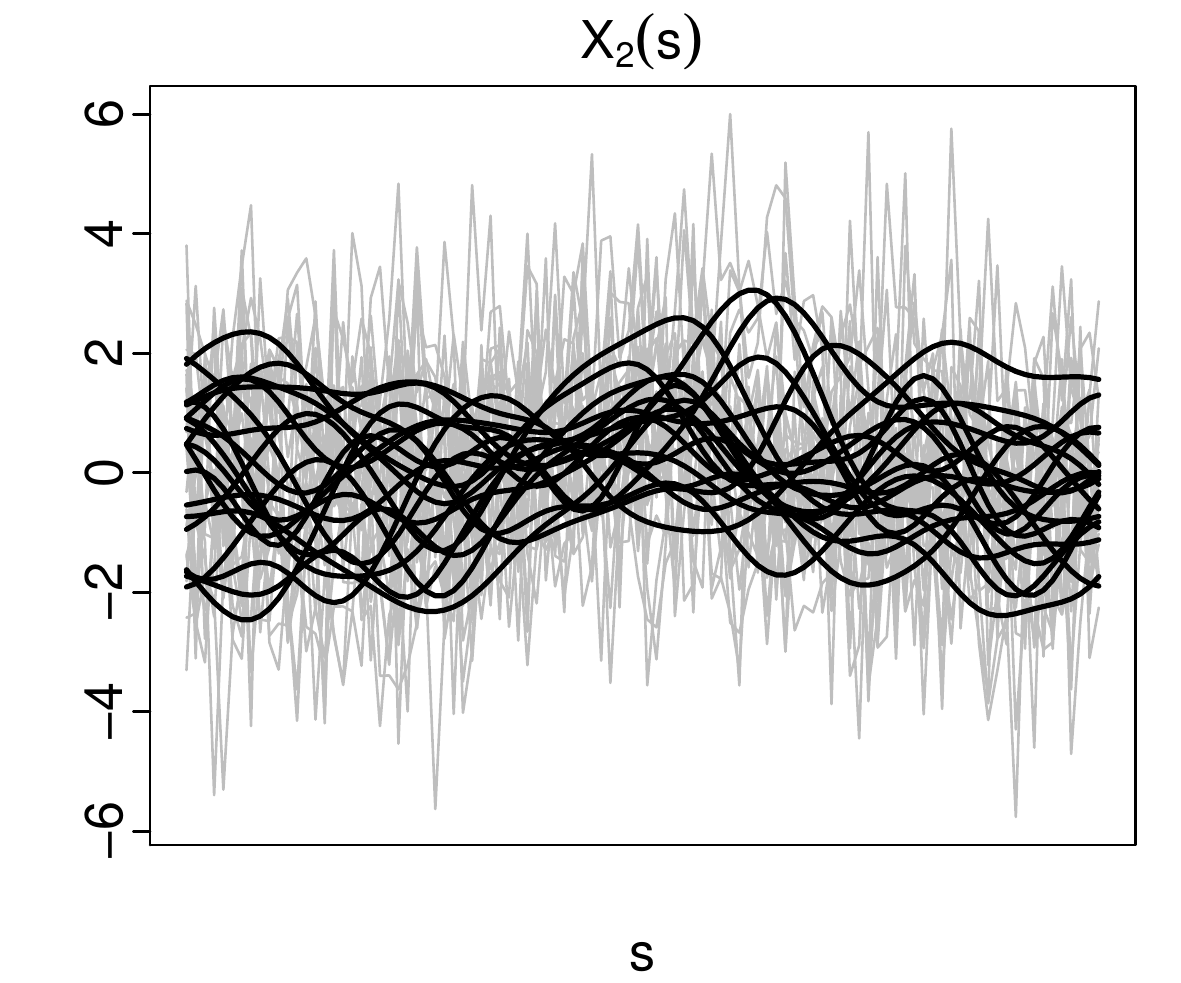}
  \includegraphics[width=5.9cm]{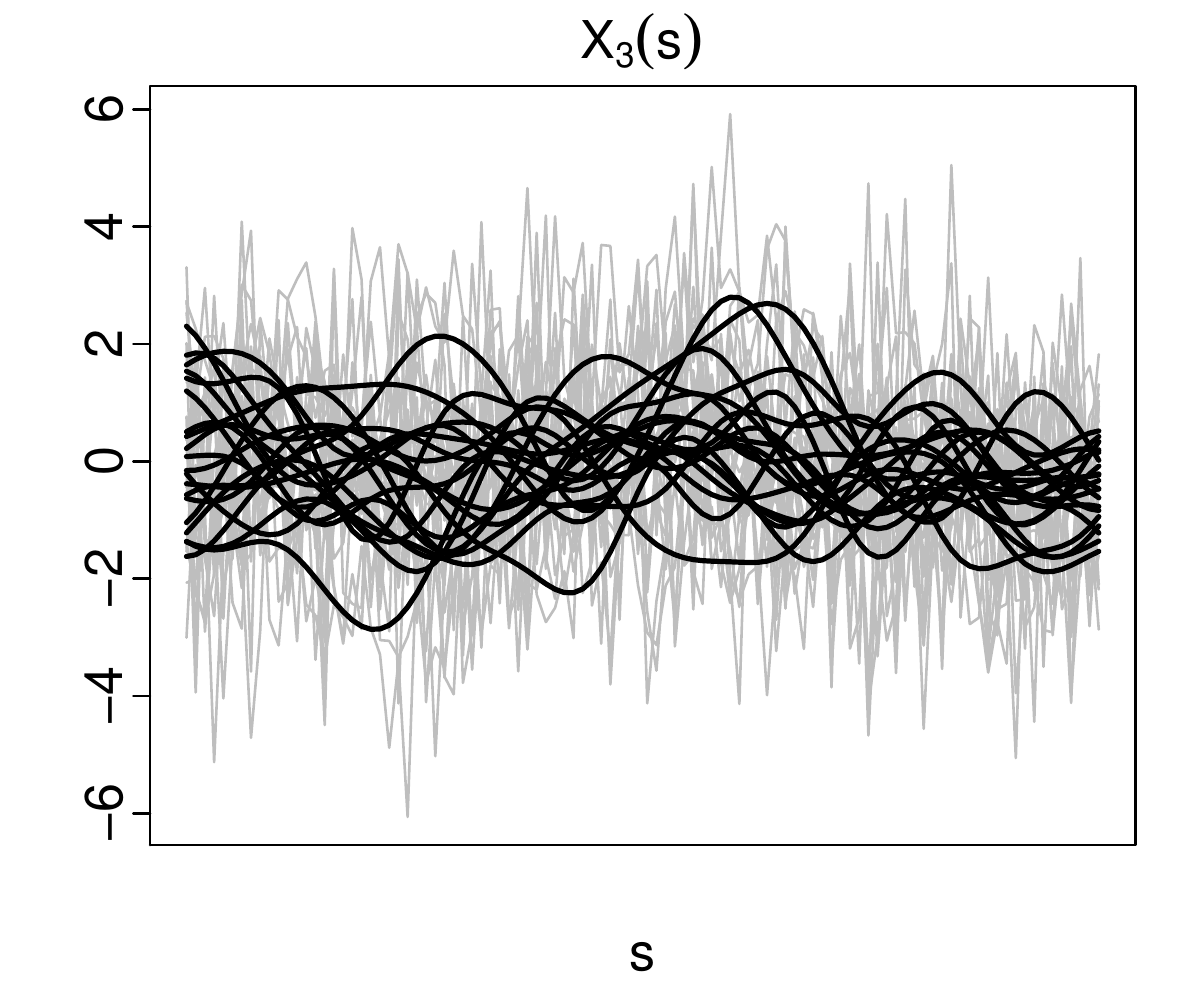}
\\  
  \includegraphics[width=5.9cm]{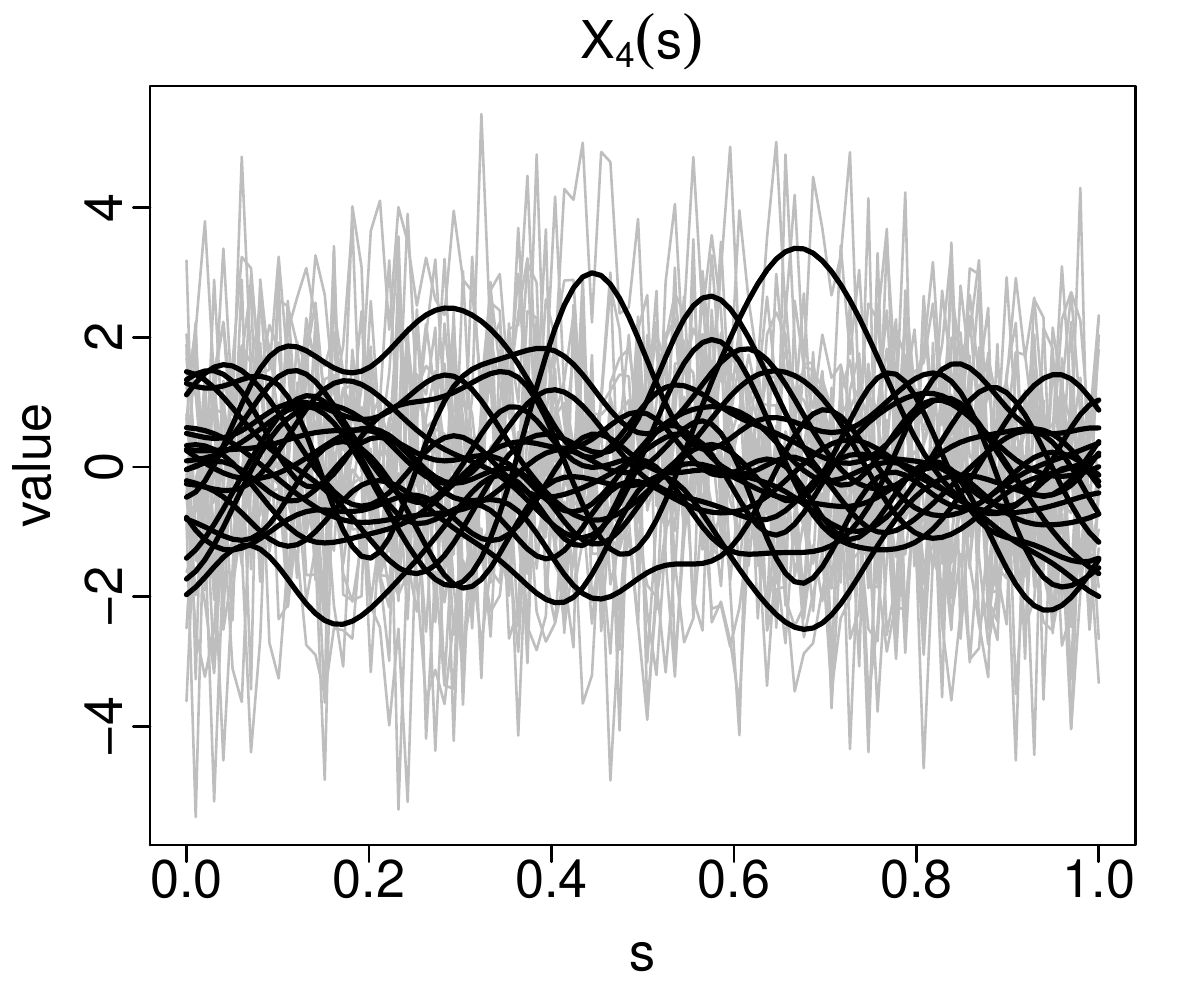}
    \includegraphics[width=5.9cm]{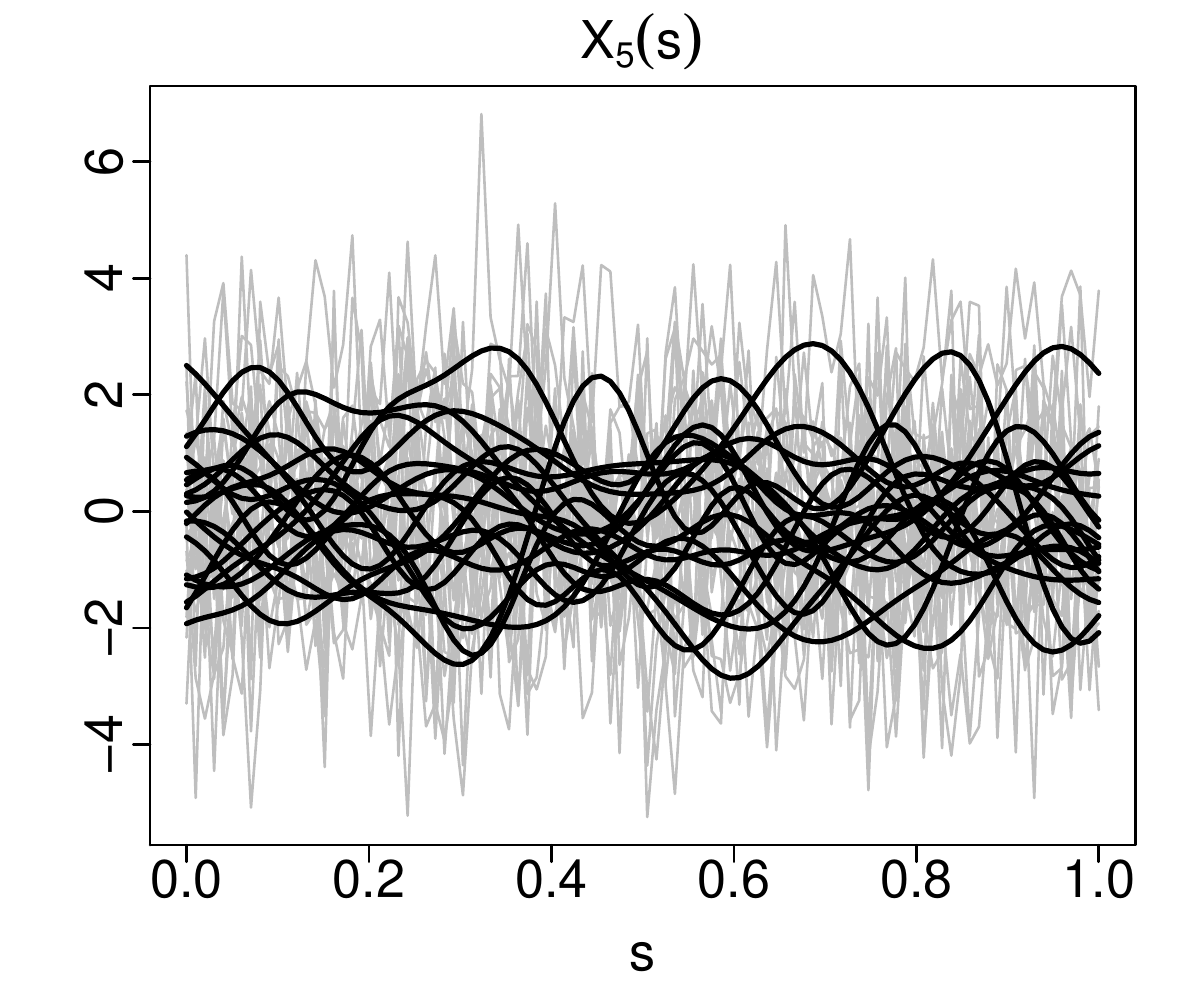}
  \includegraphics[width=5.9cm]{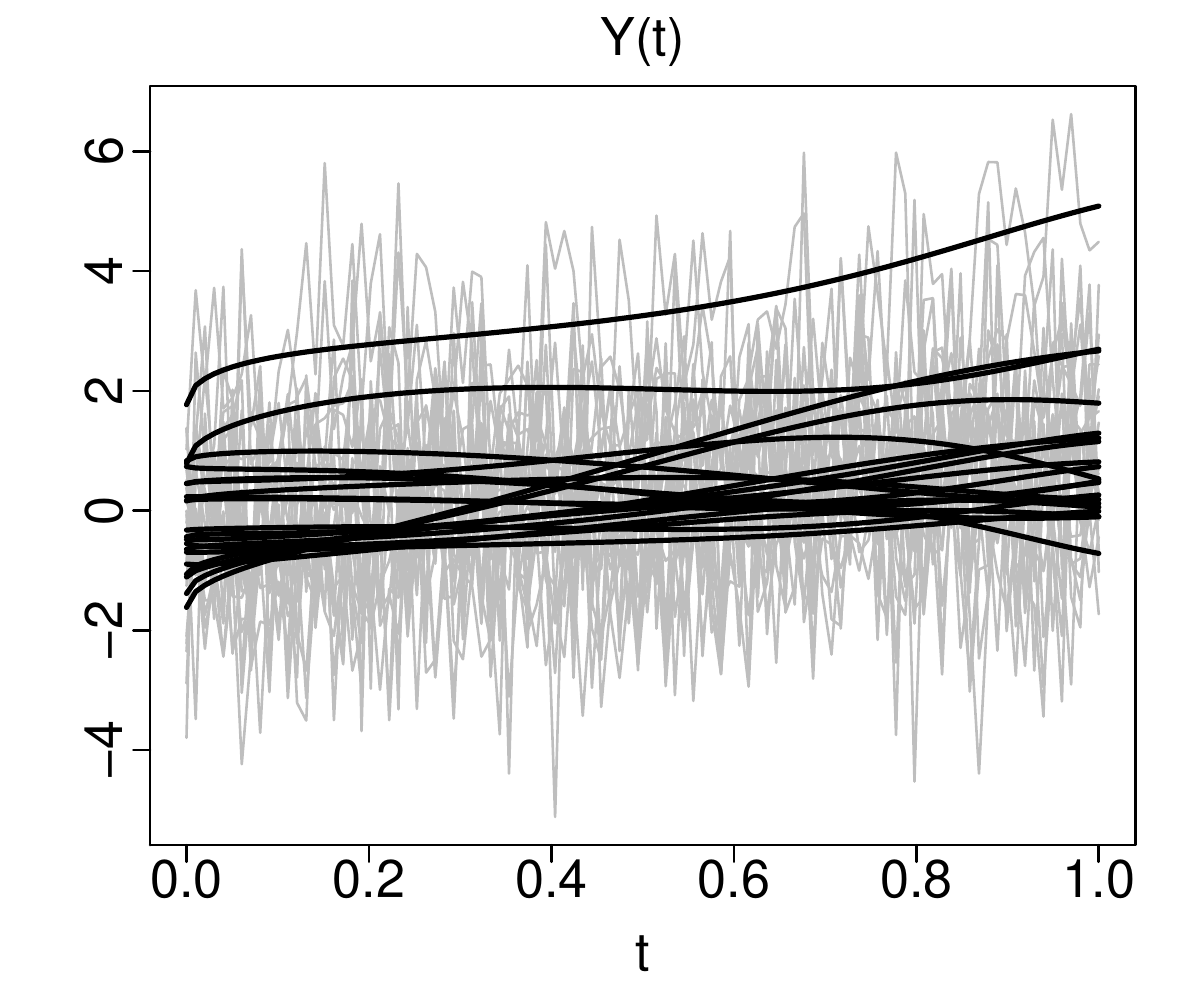}
  \caption{Plots of the generated 20 true (black lines) and noisy (gray lines) functions when $\text{Lag = 4}$. The functions were generated under the second simulation setting.}
  \label{fig:Fig_1}
\end{figure}

Several Monte-Carlo simulations were conducted to investigate the finite sample performances of the proposed method, and the results were compared with the method proposed by \cite{LuoQi} ("LQ", hereafter). Two different simulations settings, which are modified versions proposed by \cite{LuoQi}, were considered. Throughout the simulations, $m = 5$ functional predictors $\X_m$ were generated from the following process:
\begin{equation*}
\X_m(s) = \sum_{i=0}^{\text{Lag}} V_{m+i}(s) / \sqrt{\text{Lag}+1},
\end{equation*}
where $V_m(s)\text{s}~ (m = 1, \cdots, 9)$ were generated from the Gaussian process with mean zero and a positive definite covariance function $\pmb{\Sigma}_V(s,s^{\prime}) = e^{-100 (s - s^{\prime})^2}$. Herein, the parameter Lag ($\text{Lag} > 0$) controls the correlation between the predictor functions, and its large values correspond to high correlations between the predictor functions. Two different correlation levels, namely $\text{Lag} = 2$ and $\text{Lag} = 4$, were considered. The error process $\epsilon(t)$ was generated from the normal distribution with mean zero and variance four; $N(0,4)$. Then, the realizations of the response function were generated as follows:
\begin{equation*}
\Y(t) = \sum_{m \in \mathcal{M}} \int_0^1 \X_m(s) \beta_m(s,t) ds + \sum_{m,n \in \mathcal{I}} \int_0^1 \int_0^1 \X_m(s) \X_n(r) \gamma_{mn}(s,r,t) ds dr + \epsilon(t),
\end{equation*}
where $\mathcal{M} = \left\lbrace 2, 3, 4, 5 \right\rbrace$ and $\mathcal{I} = \left\lbrace  (2,2), (3,4), (4,5) \right\rbrace$ for the first simulation setting and $\mathcal{M} = \left\lbrace 1, 2, 4, 5 \right\rbrace$ and $\mathcal{I} = \left\lbrace  (1,1), (1,2), (1,5), (2,4), (4,5), (5,5) \right\rbrace$ for the second simulation setting. For the first simulation setting, the following coefficient functions were considered to generate the response:
\begin{align*}
\beta_2(s,t) &= e^{-3 (s-1)^2 -5 (t - 0.5)^2}, \\ 
\beta_3(s,t) &= e^{-5 (s-0.5)^2 - 5(t-0.5)^2} + 8 e^{-5(s-1.5)^2 - 5(t-0.5)^2}, \\
\beta_4(s,t) &= \sin(1.5 \pi s) \sin(\pi t), \\ 
\beta_5(s,t) &= \sqrt{st}, 
\end{align*}
\begin{align*}
\gamma_{22}(s,r,t) &= 5 s r \sqrt{t}, \\
\gamma_{34}(s,r,t) &= 5 \cos(\pi s) \sin(2 \pi r) \cos(2 \pi t), \\ 
\gamma_{45} &= 0.5 e^{s + 2 r -t}.
\end{align*}
For the second simulation setting, the response functions were generated using the following coefficient functions:
\begin{align*}
\beta_1(s,t) &= (s-2t)^2 / 3 , \\ 
\beta_2(s,t) &= 2 \left( \ln(1+s)\right)^2 \sin\left(2 \pi (t -0.5) \right), \\
\beta_4(s,t) &= \left( \cos \left( 1-s \right) + \sqrt{t} \right) / 3, \\ 
\beta_5(s,t) &= (1+s)^2 / \left( 3 (1 + t^2) \right) , \\
\gamma_{11}(s,r,t) &= 2(s+r)t^2, \\ 
\gamma_{1,2}(s,r,t) &= 0.01(s^2 - r^3 + t), \\ 
\gamma_{15}(s,r,t) &= 0.01 e^{2s -r + 3t}, \\
\gamma_{24}(s,r,t) &= 0.01(2s - r + 3t), \\ 
\gamma_{45}(s,r,t) &= 0.01 \frac{\ln(1 + 2s)}{1 + t}, \\ 
\gamma_{55}(s,r,t) &= \cos\left( \pi (s+r)\right) + 3 \sqrt{t}.
\end{align*}
Before fitting the regression model, the functional predictors were distorted by the Gaussian noise $u(s) \sim N(0, 4)$. All the functions for both the response and predictors variables were generated at 100 equally spaced points in the interval $[0,1]$. Examples of the true and noisy functions for the generated response and predictor variables are presented in Figure~\ref{fig:Fig_1}.

For each simulation setting, MC = 250 Monte Carlo simulations were performed, and in each simulation, $N = 300$ functions were generated for all random variables. The finite sample performances of the methods were evaluated using three metrics: MSPE, root mean squared prediction error (RMSPE), and mean absolute prediction error (MAPE). The performance metrics were calculated using the following procedure; 
\begin{inparaenum}
\item[1)] the first $N_{train} = 100$ functions of the data were used to construct the model, and 
\item[2)] the last $N_{test} = 200$ functions of the data were used to evaluate the prediction performances of the methods based on the constructed model using the first $N_{train} = 100$ functions of the data:
\end{inparaenum}
\begin{align*}
\text{MSPE} &= \frac{1}{\text{MC}} \frac{1}{N_{test}} \sum_{i = 1}^{\text{MC}} \sum_{j=1}^{N_{test}} \left\Vert \Y_{j,i}^{\text{pred}}(t) - \Y_{j,i}(t) \right\Vert^2_{\mathcal{L}_2}, \\
\text{RMSPE} &= \sqrt{\frac{1}{\text{MC}} \frac{1}{N_{test}} \sum_{i = 1}^{\text{MC}} \sum_{j=1}^{N_{test}} \left\Vert \frac{\Y_{j,i}^{\text{pred}}(t) - \Y_{j,i}(t)}{\Y_{j,i}(t)} \right\Vert^2_{\mathcal{L}_2}}, \\
\text{MAPE} &= \frac{1}{\text{MC}} \frac{1}{N_{test}} \sum_{i = 1}^{\text{MC}} \sum_{j=1}^{N_{test}} \left\Vert \frac{\left\vert \Y_{j,i}^{\text{pred}}(t) - \Y_{j,i}(t) \right\vert}{\Y_{j,i}(t)} \right\Vert_{\mathcal{L}_2},
\end{align*}
where $\Y_{j,i}(t)$ and $\Y_{j,i}^{\text{pred}}(t)$ are the generated true response functions for $j^{\text{th}}$ individual in $i^{\text{th}}$ simulation and its prediction, respectively. The proposed PLS and LQ methods were compared based on four models;
\begin{inparaenum}
\item[1)] the \textit{full model}, which includes all main, quadratic, and interaction terms;
\item[2)] the \textit{true model}, which includes only the terms used in the data-generating process;
\item[3)] the \textit{selected model}, which includes the terms selected by the variable selection procedure; and
\item[4)] the \textit{main effect model}, which includes only all five main effect terms.
\end{inparaenum}

\begin{figure}[!htbp]
  \centering
  \includegraphics[width=8.6cm]{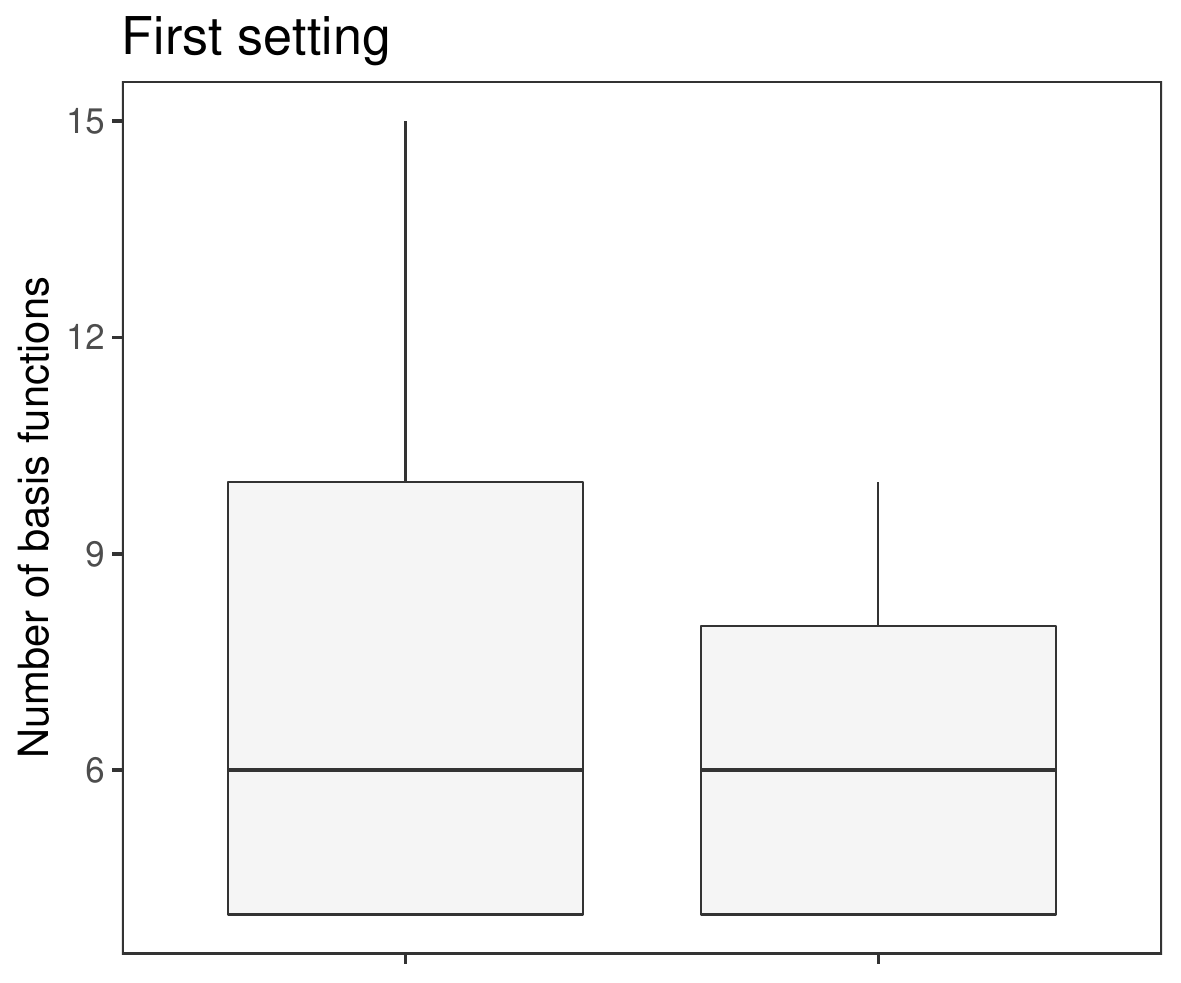}
  \qquad
  \includegraphics[width=8.6cm]{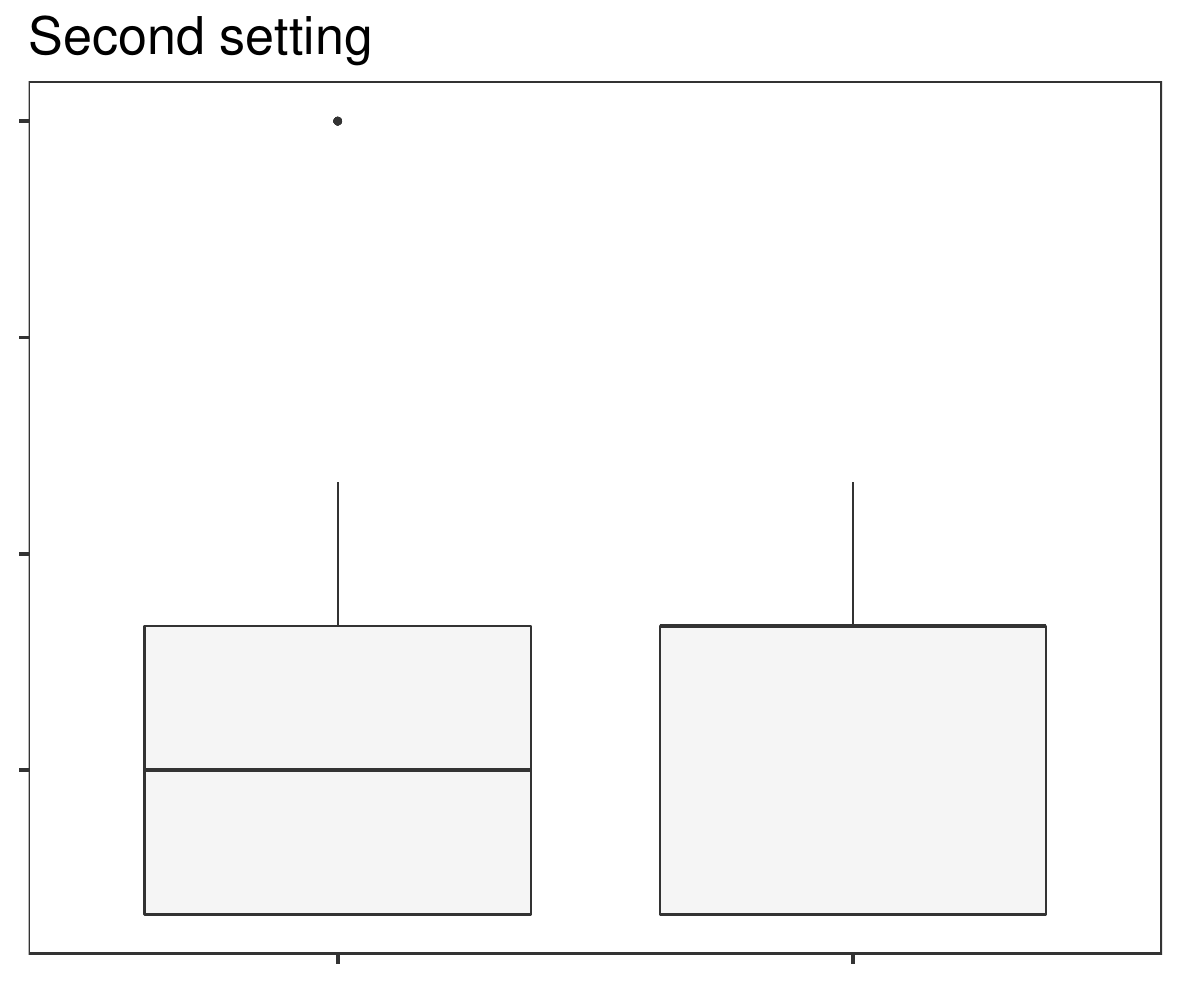}
\\  
  \includegraphics[width=8.6cm]{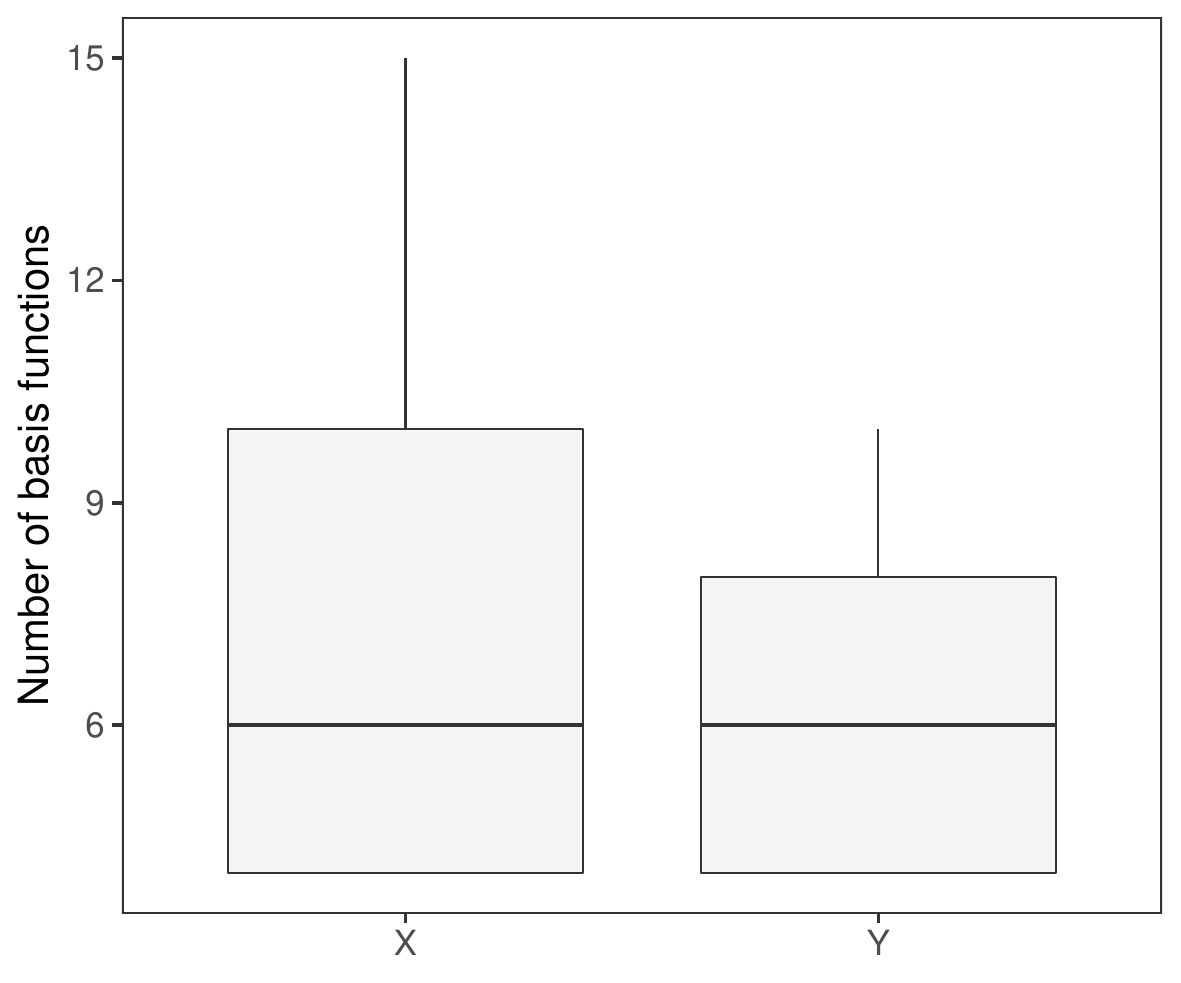}
\qquad
    \includegraphics[width=8.6cm]{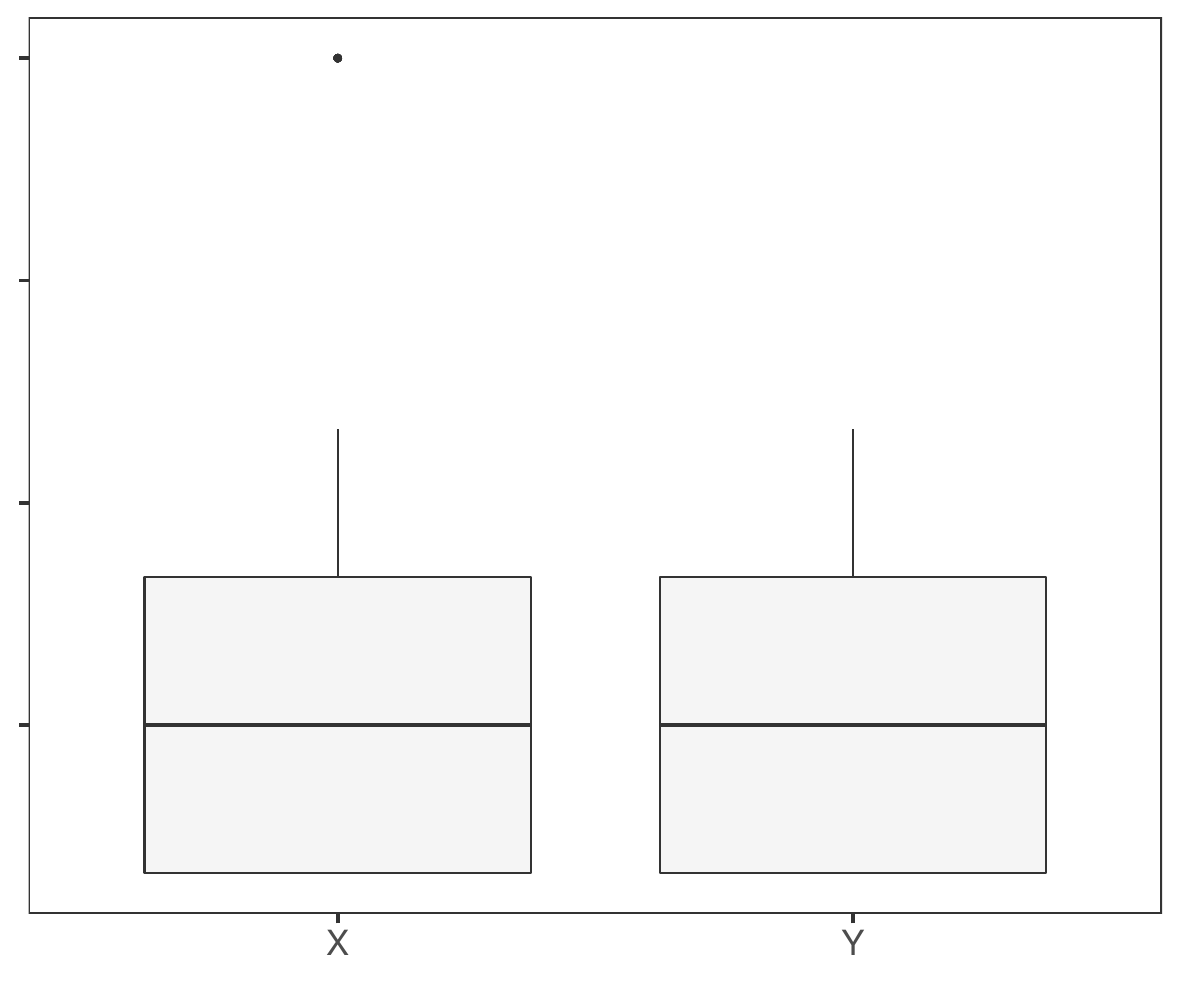}
  \caption{Boxplots of the selected numbers of basis functions for the predictor (X) and response (Y) variables from 250 Monte Carlo simulations. The first and second columns, respectively, correspond to the results obtained under the first and second simulation settings while the first and second rows correspond to the case when the correlation parameter equals Lag = 2 and Lag = 4, respectively.}
  \label{fig:Fig_2}
\end{figure}

Before presenting our results, we note that \cite{LuoQi} used 40 $B$-spline and $20 \times 20 = 400$ tensor product $B$-spline basis functions for estimating the model. For the proposed method, the number of basis functions used to approximate the response and predictor variables was chosen from a small set number of basis functions according to its prediction performance. For this purpose, first, $K_{\Y} = \left[ 4, 6, 8, 10 \right]$ and $K_{\pmb{\X}} = \left[ 4, 6, 8, 10, 15 \right]$ number of basis functions were used to approximate the response and predictor variables, respectively. Then, the number of basis functions which provide the smalles MSEs were chosen to be used in the analyses. For the proposed method, the boxplots of the selected number of basis functions for both the response and predictor variables are presented in Figure~\ref{fig:Fig_2}.

Our findings for the MSPE, RMSPE, and MAPE metrics are presented in Table~\ref{tab:pm_all}. The results demonstrate that: 
\begin{inparaenum}
\item[1)] The correlation parameter Lag has slight effects on the error measures.
\item[2)] Compared with the main effect models, both LQ and the proposed PLS methods generally produce better prediction performances (for all performance metrics) when quadratic/interaction effects are used in the model.
\item[3)] For the MSPE metric, both the proposed and LQ methods produce competitive results under the first simulation setting, while LQ outperforms the proposed PLS method when the data are generated under the second simulation setting. For the RMSPE and MAPE metrics, the proposed method shows better performance than that of LQ for both simulation settings.
\item[4)] Table~\ref{tab:pm_all} presents 12 different results (two simulation settings $\times$ two lag parameters $\times$ three error measures) for each specific model (main effect, full, true, and selected). We observe that the selected LQ model has better prediction performances than those of the full model in 58\% of all results. In contrast, the selected proposed method performs better than the full model in 91\% of all results.
\end{inparaenum}

\begin{center}
\tabcolsep 0.26in
\begin{small}
\begin{longtable}{@{}lllrrr@{}} 
\caption{Computed average MSPE, RMSPE, and MAPE values of the LQ and PLS methods. The subscripts $_{\text{main}}$, $_{\text{true}}$, $_{\text{full}}$, and $_{\text{selected}}$ correspond to the case where the model is estimated using the main effect, true, full, and selected models. The values given in brackets are the estimated standard errors for the calculated performance metrics. The smallest errors are highlighted in bold.}\\
\toprule
Setting & Lag & {Method} & {MSPE} & {RMSPE} & {MAPE} \\
\midrule
Setting 1& Lag = 2 	& LQ$_{\text{main}}$ 				& 0.428 (0.086)	& 180.353	(326.525)	& 4.519	(2.437) \\
	 	& 		 	& LQ$_{\text{full}}$ 				& 0.270 (0.067)	& 190.236	(523.532)	& 3.458	(3.807) \\
		&			& LQ$_{\text{true}}$				& \textbf{0.234} (0.045) & 151.666	(274.335)	& 3.531 (1.424) \\
		&			& LQ$_{\text{selected}}$			& 0.281 (0.051) & 197.873	(634.627)	& 3.370 (4.571) \\
		&			& PLS$_{\text{main}}$ 				& 0.433 (0.087)	& 80.861	(150.235)	& 3.366	(1.051) \\
	 	& 		 	& PLS$_{\text{full}}$ 				& 0.311 (0.049)	& 96.755	(284.596)	& 2.376	(1.545) \\
		&			& PLS$_{\text{true}}$				& 0.238 (0.036) & 67.417	(189.168)	& \textbf{1.799} (1.053) \\
		&			& PLS$_{\text{selected}}$			& 0.244 (0.048) & \textbf{64.563}	(130.164)	& 1.926 (1.015) \\
\midrule
	 	& Lag = 4 	& LQ$_{\text{main}}$ 				& 0.522 (0.128)	& 229.012	(354.072)	& 5.980	(2.809) \\
	 	& 		 	& LQ$_{\text{full}}$ 				& 0.284 (0.055)	& 142.713	(266.679)	& 3.194	(1.449) \\
		&			& LQ$_{\text{true}}$				& 0.262 (0.049) & 126.574	(239.987)	& 2.833 (1.410) \\
		&			& LQ$_{\text{selected}}$			& 0.298 (0.061) & 114.923	(220.162)	& 2.912 (1.098) \\
		&			& PLS$_{\text{main}}$ 				& 0.527 (0.124)	& 136.552	(217.661)	& 3.871	(1.520) \\
	 	& 		 	& PLS$_{\text{full}}$ 				& 0.288 (0.054)	& 45.590	(41.800)	& 2.086	(0.992) \\
		&			& PLS$_{\text{true}}$				& \textbf{0.258} (0.042) & 40.495	(38.179)	& \textbf{1.889} (0.889) \\
		&			& PLS$_{\text{selected}}$			& 0.260 (0.045) & \textbf{40.406}	(41.207)	& 2.142 (1.371) \\
\bottomrule
Setting 2& Lag = 2 	& LQ$_{\text{main}}$ 				& 0.273 (0.059)	& 256.474	(392.914)	& 6.065	(3.726) \\
	 	& 		 	& LQ$_{\text{full}}$ 				& 0.115 (0.035)	& 139.166	(224.285)	& 2.582	(1.616) \\
		&			& LQ$_{\text{true}}$				& \textbf{0.101} (0.027) & 146.398	(230.747)	& \textbf{2.341} (1.104) \\
		&			& LQ$_{\text{selected}}$			& 0.103 (0.028) & 158.856	(382.274)	& 2.932 (3.491) \\
		&			& PLS$_{\text{main}}$ 				& 0.255 (0.054)	& 169.522	(250.788)	& 4.877	(2.072) \\
	 	& 		 	& PLS$_{\text{full}}$ 				& 0.157 (0.029)	& 89.768	(192.956)	& 2.847	(1.058) \\
		&			& PLS$_{\text{true}}$				& 0.140 (0.028) & 89.620	(195.483)	& 2.724 (1.027) \\
		&			& PLS$_{\text{selected}}$			& 0.118 (0.027) & \textbf{72.743}	(137.590)	& 2.452 (1.011) \\
\midrule
	 	& Lag = 4 	& LQ$_{\text{main}}$ 				& 0.284 (0.075)	& 282.228	(490.338)	& 6.713	(4.571) \\
	 	& 		 	& LQ$_{\text{full}}$ 				& 0.129 (0.028)	& 160.912	(264.747)	& 3.385	(1.919) \\
		&			& LQ$_{\text{true}}$				& 0.111 (0.027) & 138.492	(222.025)	& 3.001 (1.258) \\
		&			& LQ$_{\text{selected}}$			& \textbf{0.109} (0.024) & 157.069	(268.081)	& 3.218 (1.769) \\
		&			& PLS$_{\text{main}}$ 				& 0.267 (0.069)	& 157.322	(280.071)	& 5.045	(2.326) \\
	 	& 		 	& PLS$_{\text{full}}$ 				& 0.154 (0.040)	& 98.047	(168.614)	& 3.174	(1.950) \\
		&			& PLS$_{\text{true}}$				& 0.136 (0.029) & 71.935	(106.868)	& 2.751 (1.263) \\
		&			& PLS$_{\text{selected}}$			& 0.131 (0.035) & \textbf{70.139}	(109.836)	& \textbf{2.747} (1.201) \\
\bottomrule
\label{tab:pm_all}
\end{longtable}
\end{small}
\end{center}

Moreover, we compared both methods in terms of their computing times. The computing time of the LQ method is mainly based on the number of one-dimensional basis functions used to construct the tensor product basis functions to estimate the functions of quadratic/interaction components. Our records showed that, compared with the LQ method using the default number of basis functions (40 $B$-spline and $20 \times 20 = 400$ tensor product $B$-spline basis functions), the proposed method requires a similar computing time when a small number of basis functions are used to estimate the model, while our proposed method requires more computing time than the LQ method as the number of basis functions used in the model increases. This is because, for the proposed method, the dimension of $\pmb{\Psi}$, $\text{dim}\left( \pmb{\Psi}\right) = K_{\pmb{\X}}^2 \times K_{\pmb{\X}}^2$, increases exponentially as the number of basis functions used in the model increases.  For example, the proposed method requires $[1.21, 4.18, 7.51]$ times more computing times than the LQ method when $K_{\Y} = K_{\pmb{\X}} = \left[8, 12, 20 \right]$ number of basis functions are used to estimate the model. In summary, our findings have demonstrated that the proposed PLS method generally results in improved accuracy at the expense of longer computational time in comparison with the LQ method.

We note that our numerical analyses were performed using \texttt{R} 3.6.0 (an example R code can be found at \url{https://github.com/UfukBeyaztas/FLM_interaction}.)

\section{Empirical data examples} \label{sec:real}

\subsection{Hawaii ocean data}

\begin{figure}[!htbp]
  \centering
  \includegraphics[width=6cm]{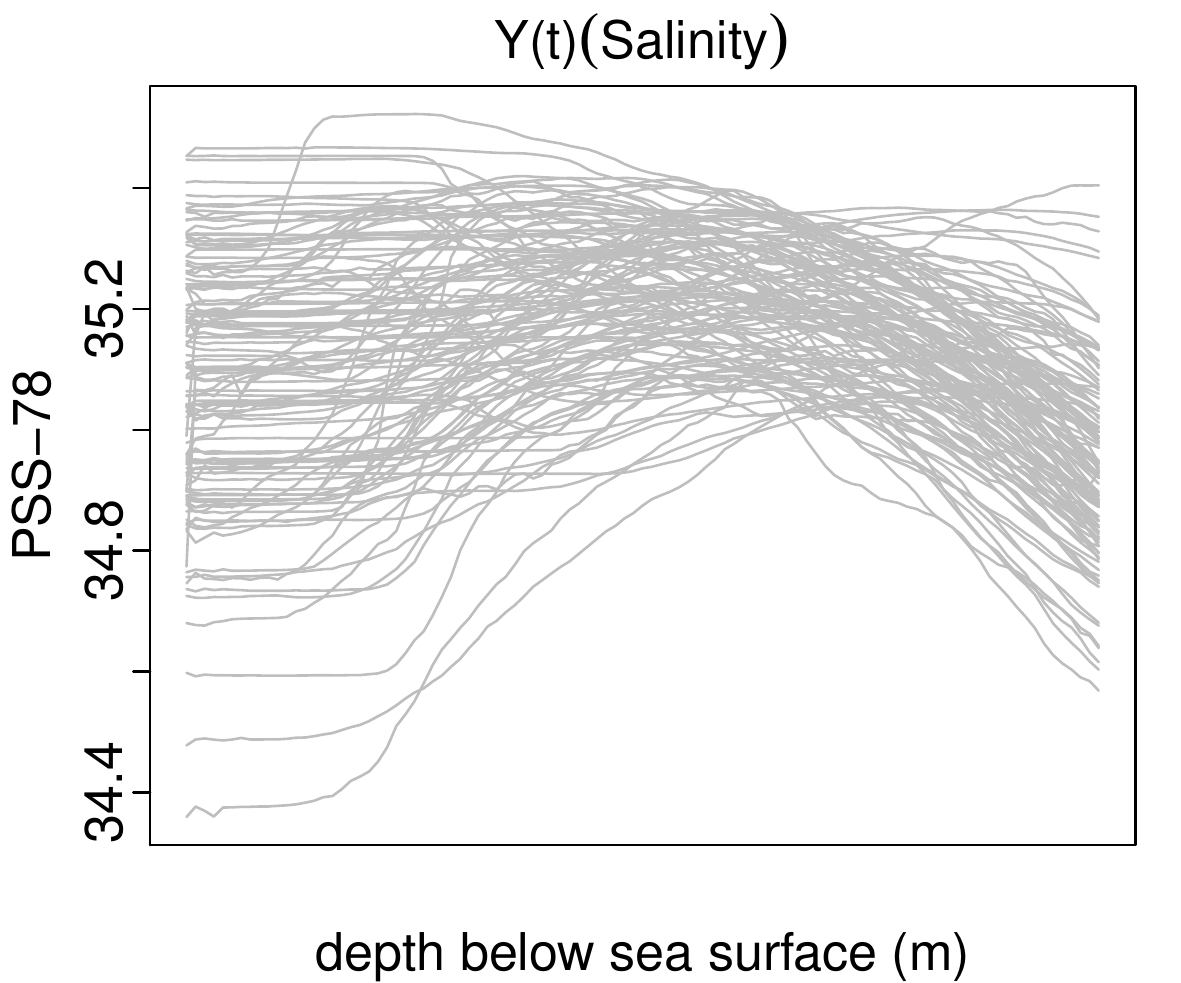}
  \includegraphics[width=6cm]{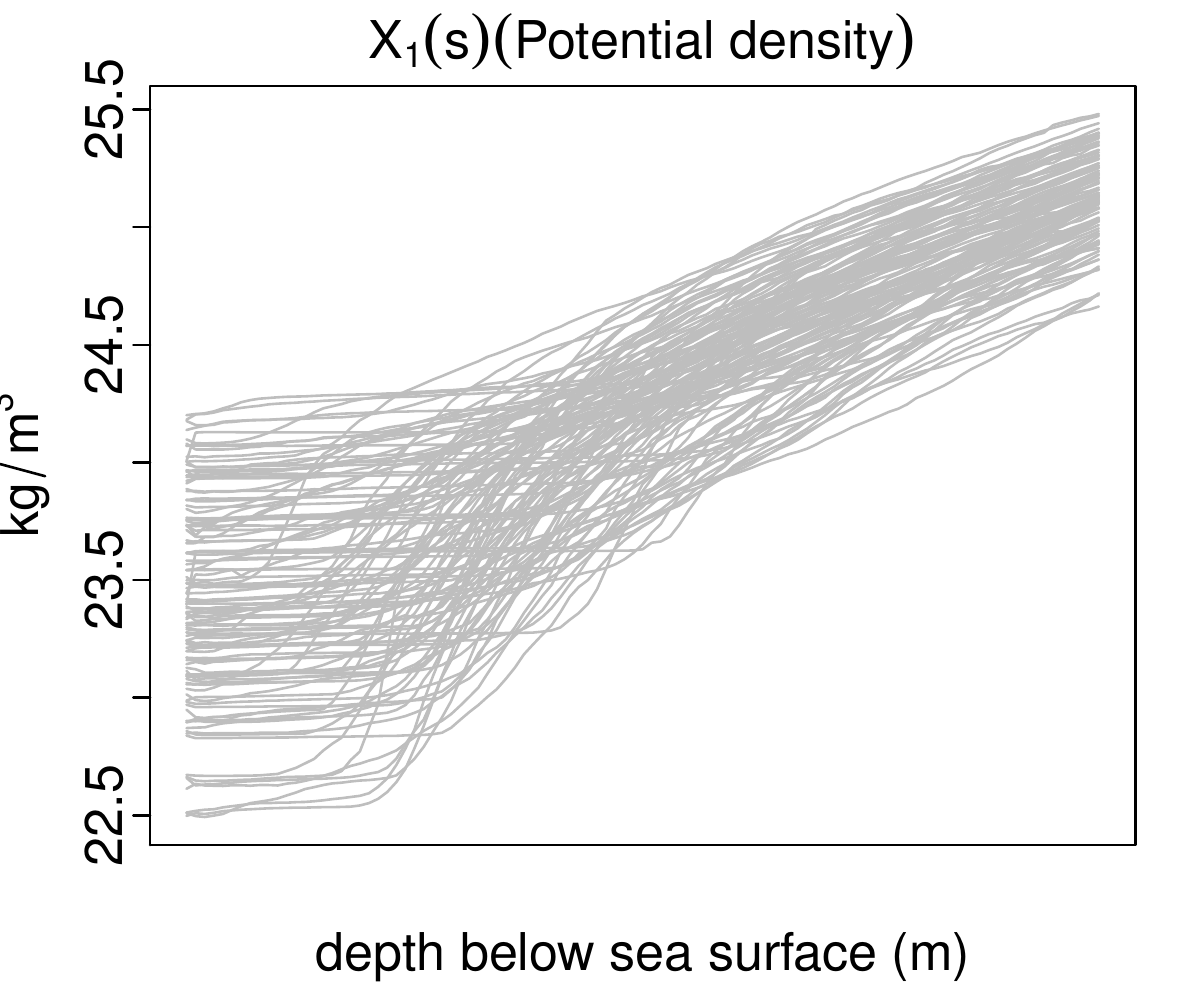}
  \includegraphics[width=6cm]{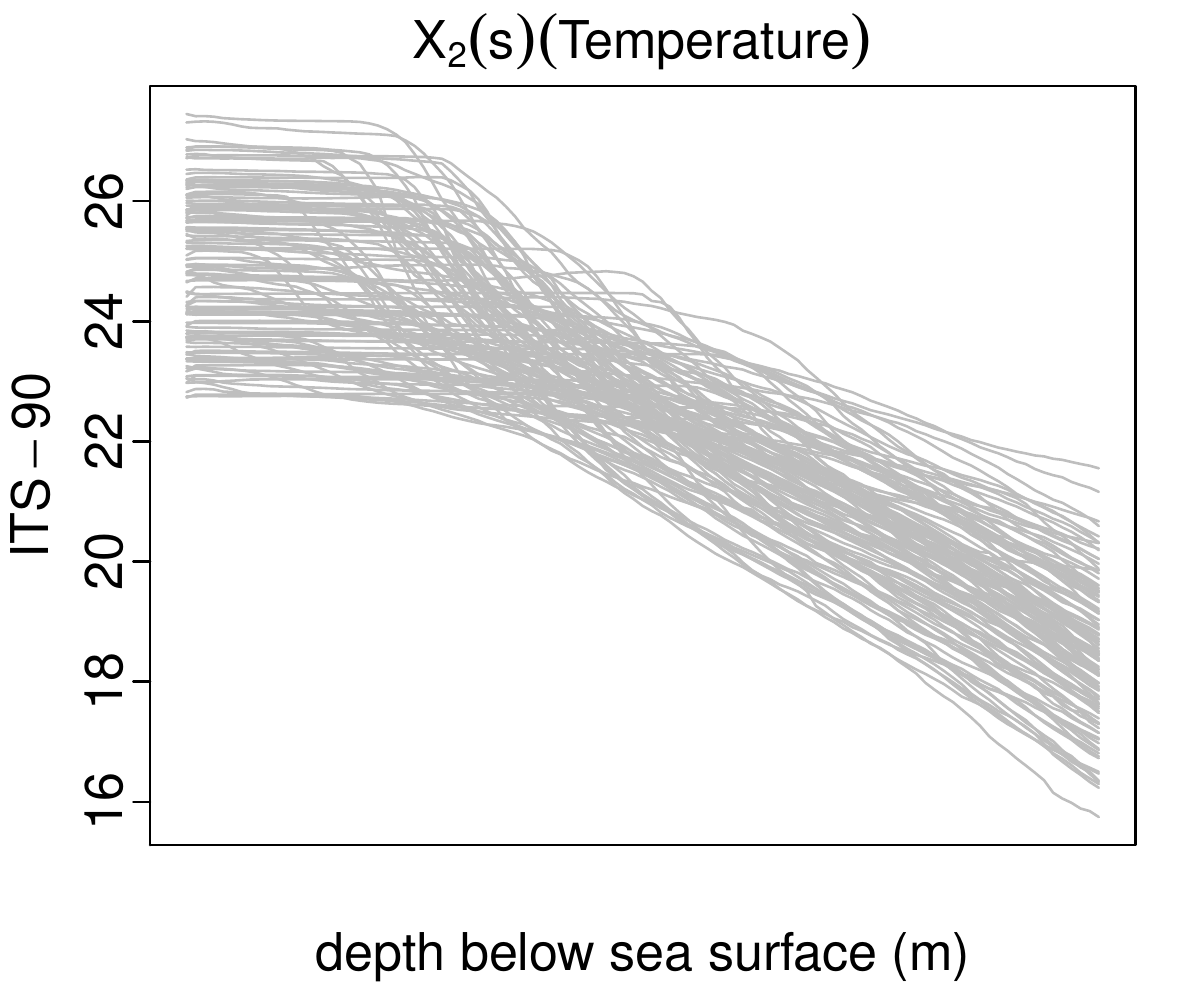}
\\  
  \includegraphics[width=6cm]{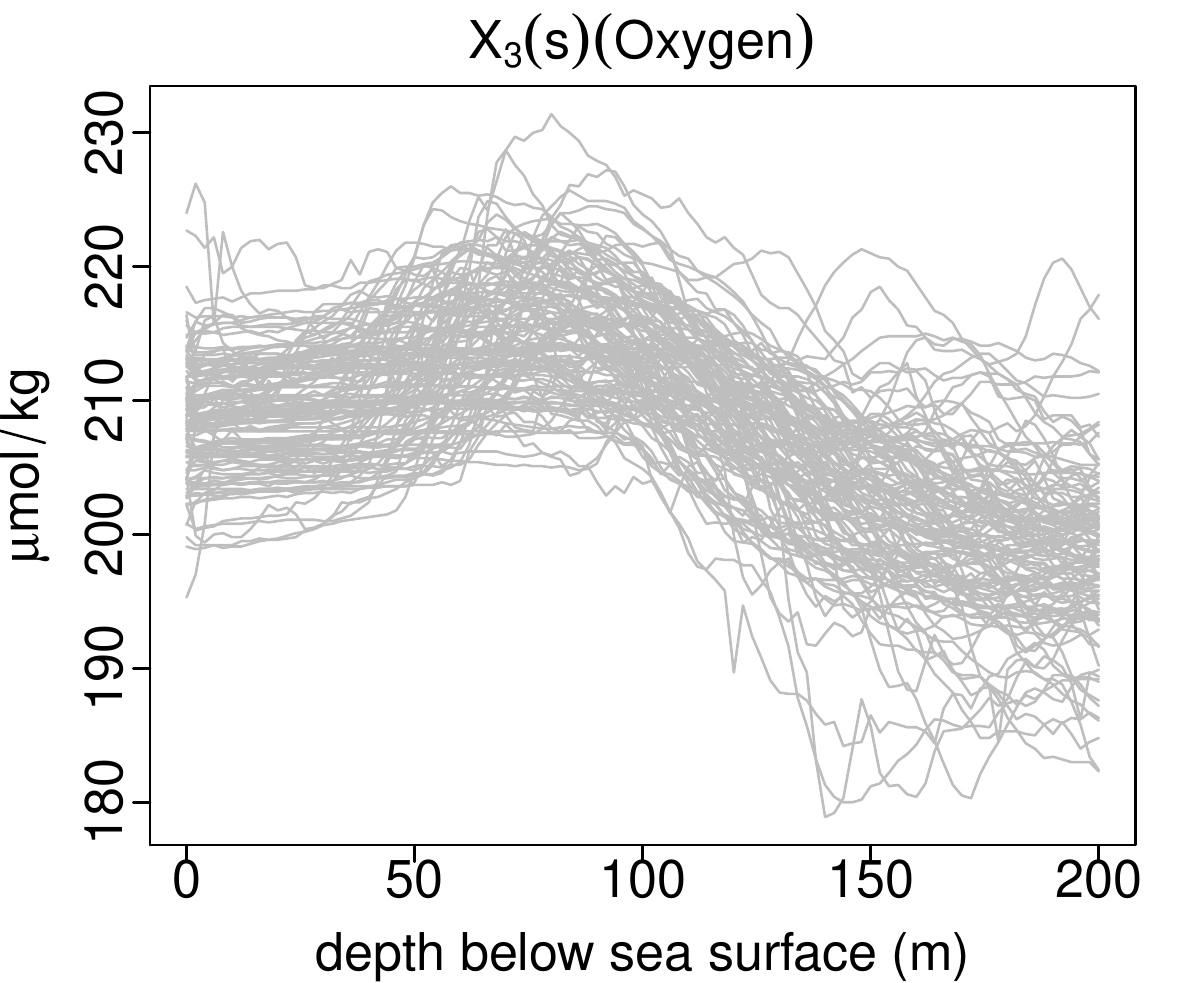}
    \includegraphics[width=6cm]{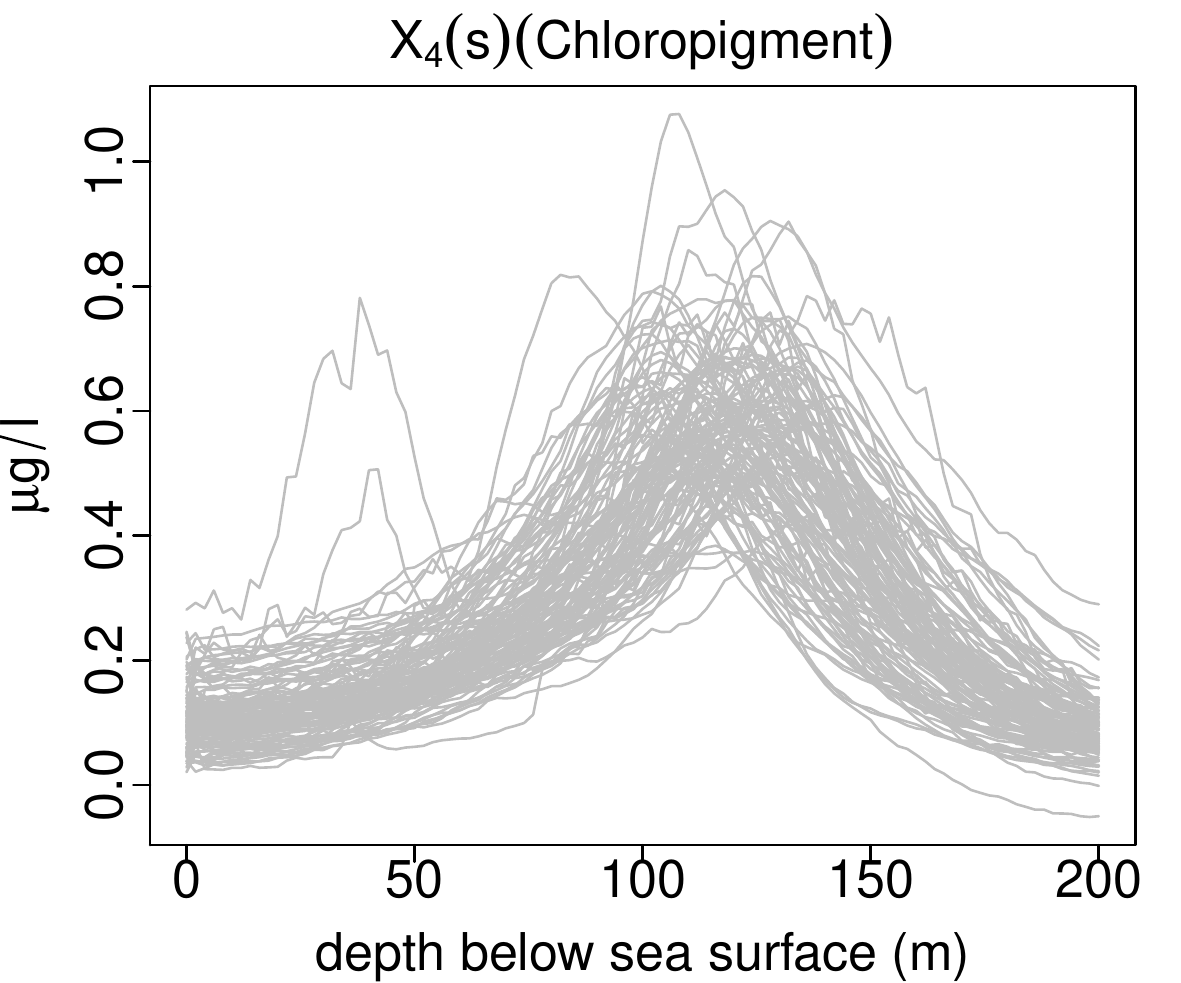}
  \caption{Plots of the functional variables: salinity, potential density, temperature, oxygen, and chloropigment in the Hawaii ocean dataset. The observations are the functions of depth below the sea surface and $0 \leq s,t \leq 200$. PSS-78 and ITS-90 denote the practical salinity scale of 1978 and the international temperature scale of 1990, respectively.}
  \label{fig:Fig_3}
\end{figure}

Available in the \texttt{R} package ``FRegSigCom'' from \cite{LuoQi_R}, the Hawaii ocean dataset consists of five functional variables: salinity, potential density, temperature, oxygen, and chloropigment. These variables are measured every 2m between 0 and 200m below the sea surface and are viewed as functions of depth \citep[see][for a more detailed description of the dataset]{LuoQi}. The functional variables in this dataset consist of 116 curves, which are observed at 101 equally spaced points in the interval $[0, 200]$. For this dataset, we considered predicting salinity using the other four variables. The plots of the functional variables are presented in Figure~\ref{fig:Fig_3}.

In the prediction process, the dataset was divided into the following training and testing samples; the model was constructed based on 90 randomly selected curves to predict the remaining 26 curves of the salinity variable. This procedure was repeated 100 times, and for each replication, the performance metrics (MSPE, RMSPE, and MAPE) were calculated for the main effect, full, and selected models. In addition, the following two coefficients of determination were calculated for each replication to measure the goodness of fit ($R^2$) and the relative predictive performances ($R^2_{pred}$) of the models \citep{LuoQi}:
\begin{equation*}
R^2 = 1 - \frac{\sum_{i=1}^{90} \left\Vert \Y^{fit}_i(t) - \Y^{train}_i(t) \right\Vert^2_{\mathcal{L}_2}}{\sum_{i=1}^{90} \left\Vert \Y^{train}_i(t) - \overline{\Y}^{train}(t) \right\Vert^2_{\mathcal{L}_2}}, \qquad
R^2_{pred} = 1 - \frac{\sum_{i=1}^{26} \left\Vert \Y^{pred}_i(t) - \Y^{test}_i(t) \right\Vert^2_{\mathcal{L}_2}}{\sum_{i=1}^{26} \left\Vert \Y^{test}_i(t) - \overline{\Y}^{test}(t) \right\Vert^2_{\mathcal{L}_2}},
\end{equation*}
where $\left\lbrace \Y^{train}_i(t), \Y^{fit}_i(t) \right\rbrace$, for $i = 1, \cdots, 90$, are respectively, the $i^{\text{th}}$ observed and fitted curves of the salinity variable in the training sample, and $\left\lbrace \Y^{test}_i(t), \Y^{pred}_i(t) \right\rbrace$, for $i = 1, \cdots, 26$, are the $i^{\text{th}}$ observed and predicted curves of the salinity variable in the testing sample, respectively. For this dataset, $K_{\Y} = K_{\pmb{\X}} = 20$ number of basis functions were used to calculate the performance metrics for the proposed PLS method.

Our findings are reported in Table~\ref{tab:pm_oc}. The results show that the selected models of both LQ and the proposed method have better performances than their full and main effect models. The coefficients of determination values produced by both methods are very close to 1, and these results demonstrate that both methods have good prediction performances. However, as shown in Table~\ref{tab:pm_oc}, the selected model of the proposed PLS method has an improved prediction performance compared with the selected model of LQ. 

\begin{table}[!htbp]
\begin{center}
\tabcolsep 0.22in
\caption{Computed average MSPE, RMSPE, MAPE, $R^2$, and $R^2_{pred}$ values of the LQ and PLS methods for the Hawaii ocean data. The subscripts $_{\text{main}}$, $_{\text{full}}$, and $_{\text{selected}}$ respectively correspond to the case where the model is estimated using the main effect, full, and selected models. The values given in brackets are the estimated standard errors for the calculated performance metrics.}\label{tab:pm_oc}
\begin{tabular}{@{}lrrrrr@{}} 
\toprule
{Method} & {MSPE} & {RMSPE} & {MAPE} & {$R^2$} & {$R^2_{pred}$} \\
\midrule
LQ$_{\text{main}}$ 			& 2.6 $\times 10^{-4}$ 	& 4.6 $\times 10^{-4}$ 	& 3.3 $\times 10^{-4}$ 	& 0.994 	& 0.999  \\
	& (7.1 $\times 10^{-5}$) & (5.8 $\times 10^{-5}$) & (3.7 $\times 10^{-5}$) & (1.3 $\times 10^{-3}$) & (4.2 $\times 10^{-5}$)\\
LQ$_{\text{full}}$			& 2.1 $\times 10^{-4}$ 	& 4.1 $\times 10^{-4}$ 	& 2.8 $\times 10^{-4}$ 	& 0.994 	& 0.999 \\
& (7.5 $\times 10^{-5}$) &(6.7 $\times 10^{-5}$) & (4.9 $\times 10^{-5}$) & (1.2 $\times 10^{-3}$) &  (3.1 $\times 10^{-5}$) \\
LQ$_{\text{selected}}$		& 2.0 $\times 10^{-4}$ 	& 3.9 $\times 10^{-4}$ 	& 2.7 $\times 10^{-4}$ 	& 0.994 	& 0.999  \\
& (7.5 $\times 10^{-5}$) & (6.8 $\times 10^{-5}$) & (4.6 $\times 10^{-5}$) & (1.2 $\times 10^{-3}$) & (3.1 $\times 10^{-5}$) \\
PLS$_{\text{main}}$ 		& 4.6 $\times 10^{-4}$ 	& 5.8 $\times 10^{-4}$ 	& 4.3 $\times 10^{-4}$ 	& 0.991 	& 0.990 \\
& (2.1 $\times 10^{-4}$) & (1.2 $\times 10^{-4}$) & (7.0 $\times 10^{-5}$) & (2.6 $\times 10^{-5}$) & (3.1 $\times 10^{-4}$)  \\
PLS$_{\text{full}}$			& 4.3 $\times 10^{-4}$ 	& 5.8 $\times 10^{-4}$ 	& 4.0$\times 10^{-4}$ 	& 0.993 	& 0.993  \\
& (2.0 $\times 10^{-4}$) & (1.2 $\times 10^{-4}$) & (4.7 $\times 10^{-5}$) & (2.6 $\times 10^{-5}$) & (5.5 $\times 10^{-4}$) \\
PLS$_{\text{selected}}$		& 1.9 $\times 10^{-5}$ 	& 1.2 $\times 10^{-4}$ 	& 8.6 $\times 10^{-5}$ 	& 0.995 	& 0.999 \\
& (5.5 $\times 10^{-6}$) & (1.7 $\times 10^{-5}$) & (8.8 $\times 10^{-6}$) & (2.2 $\times 10^{-5}$) & (1.8 $\times 10^{-5}$)  \\
\bottomrule
\end{tabular}
\end{center}
\end{table}

To determine the significant main and quadratic/interaction effects terms for this dataset, a regression model was constructed based on all 116 curves. According to our results, the potential density $\X_1(s)$ and temperature $\X_2(s)$ were selected as the significant main effects by the forward selection procedures of both the LQ and PLS methods. For the quadratic and interaction effects terms, $\X_1(s) \X_1(r)$, $\X_1(s) \X_2(r)$, and $\X_2(s) \X_2(r)$ were selected as significant by LQ, whereas $\X_1(s) \X_1(r)$ and $\X_1(s) \X_2(r)$ were selected as significant by our proposed method. 

\begin{figure}[!htbp]
  \centering
  \includegraphics[width=8.8cm]{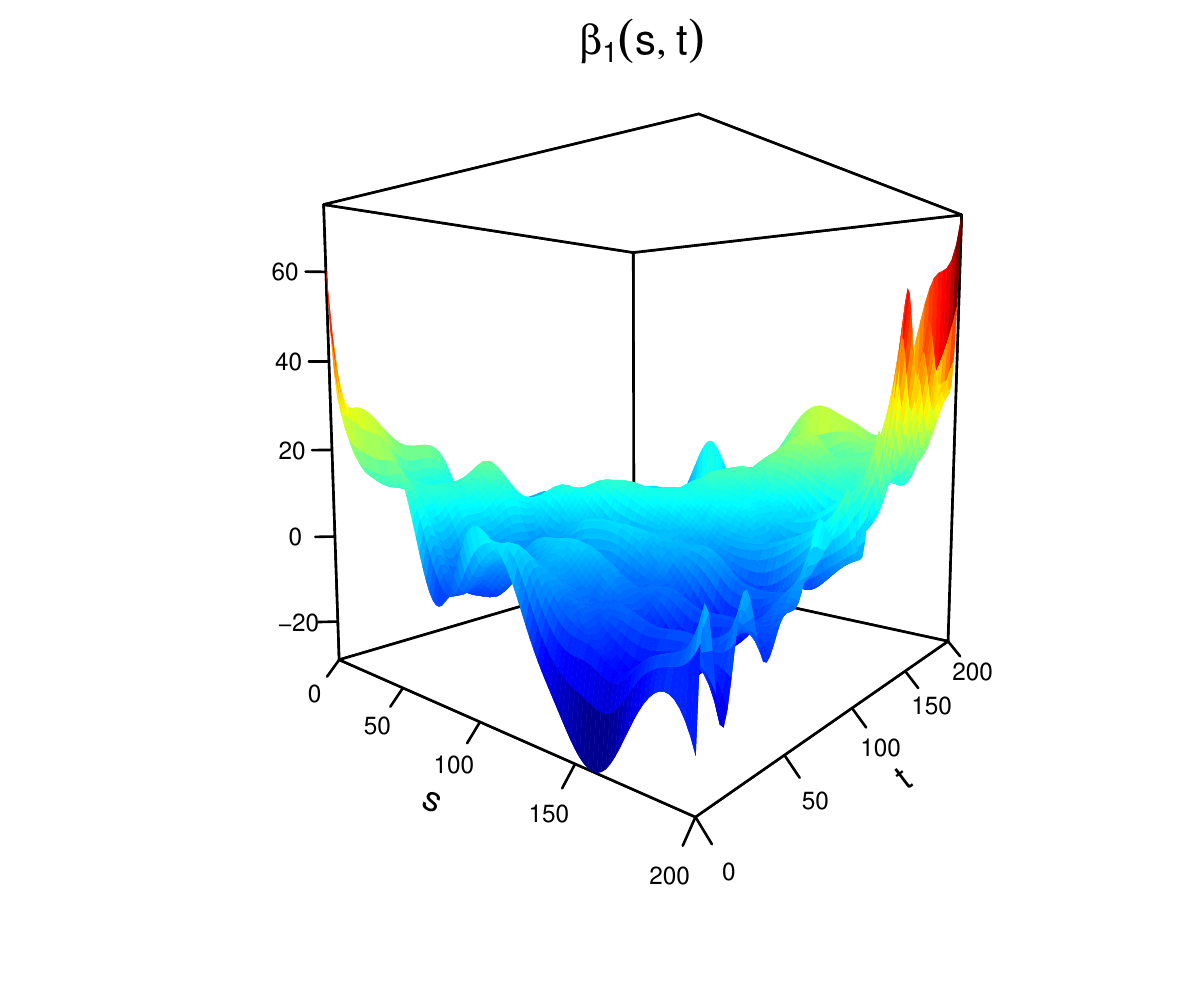}
\quad
  \includegraphics[width=8.8cm]{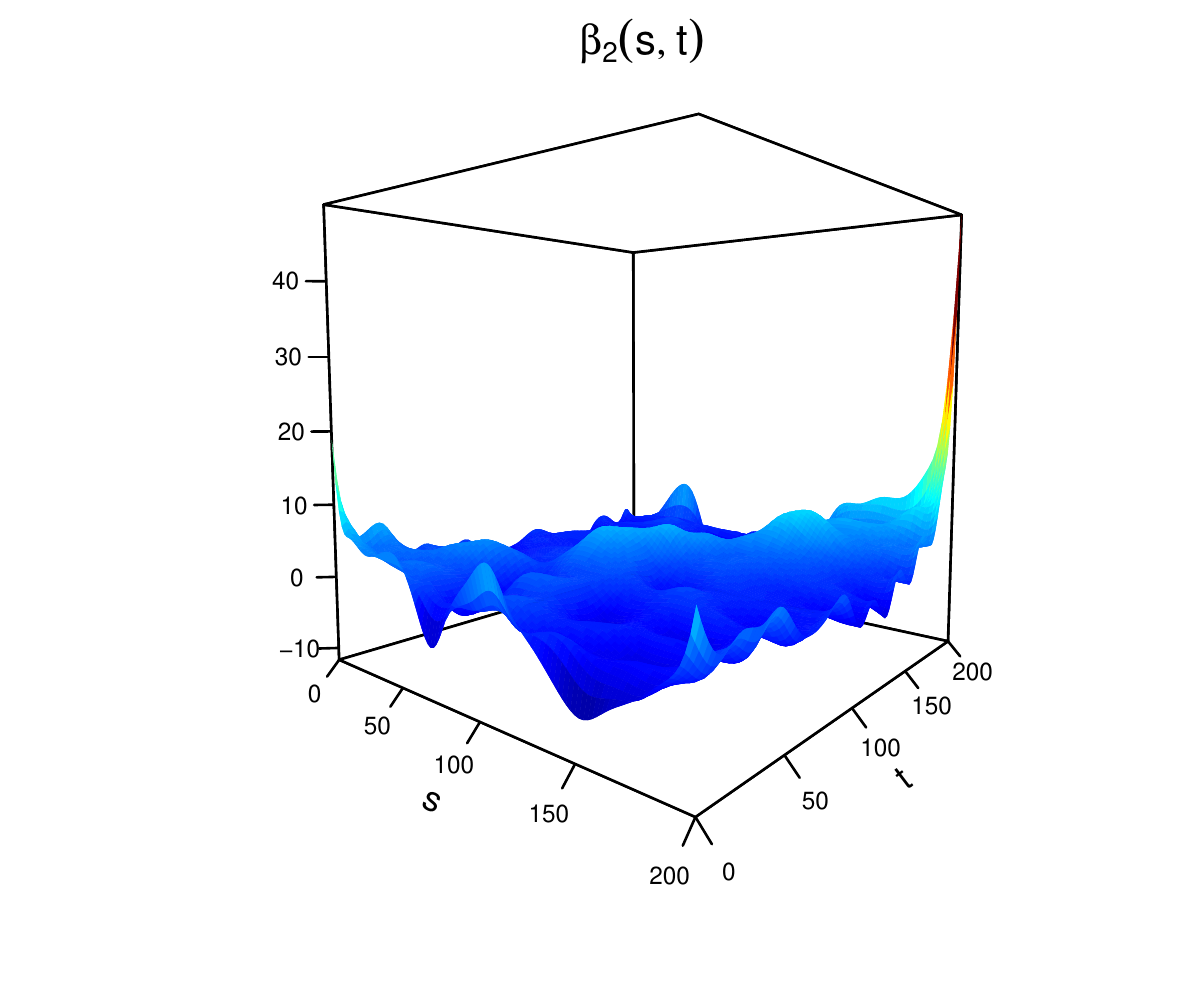}
\\  
  \includegraphics[width=8.8cm]{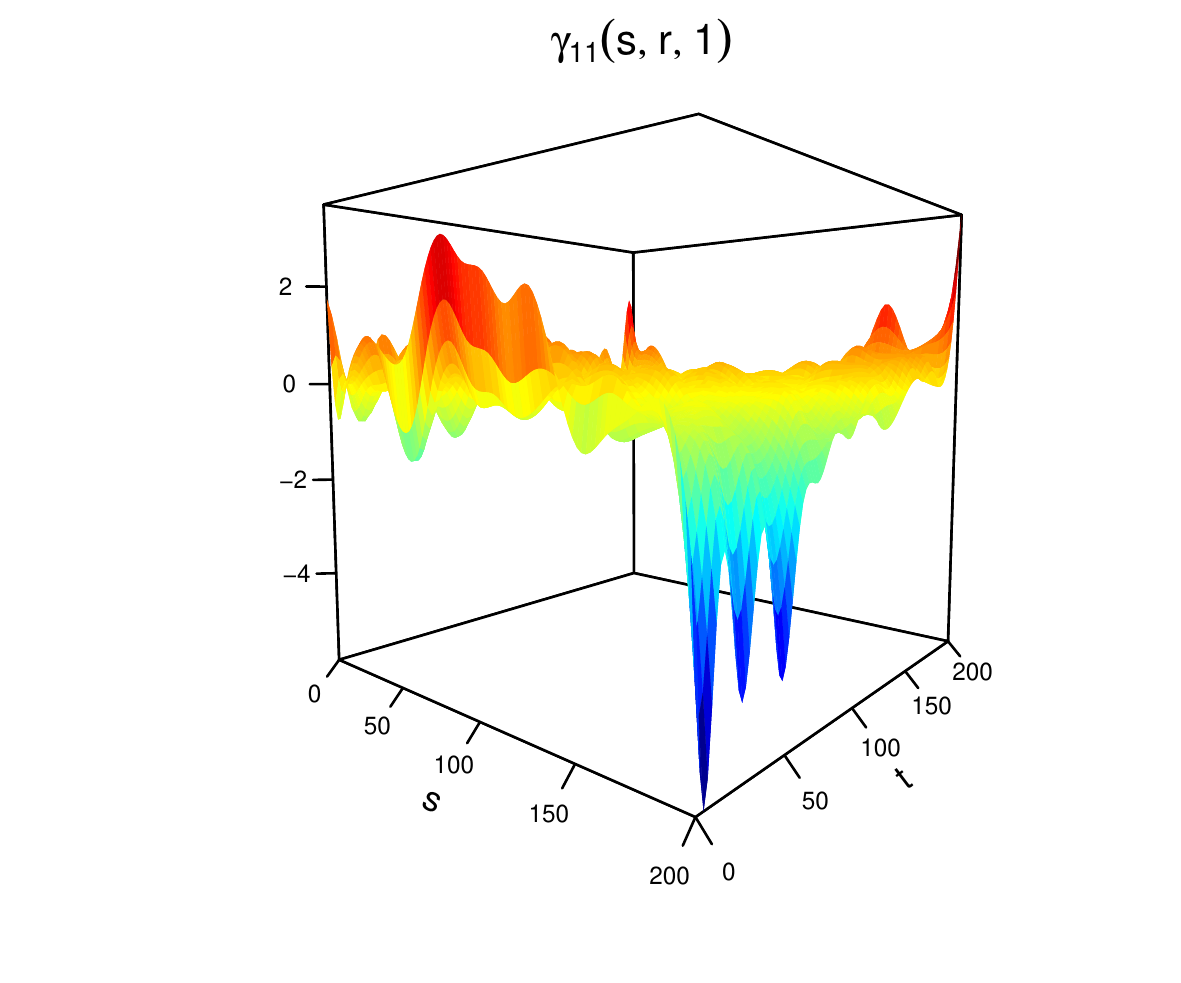}
\quad
    \includegraphics[width=8.8cm]{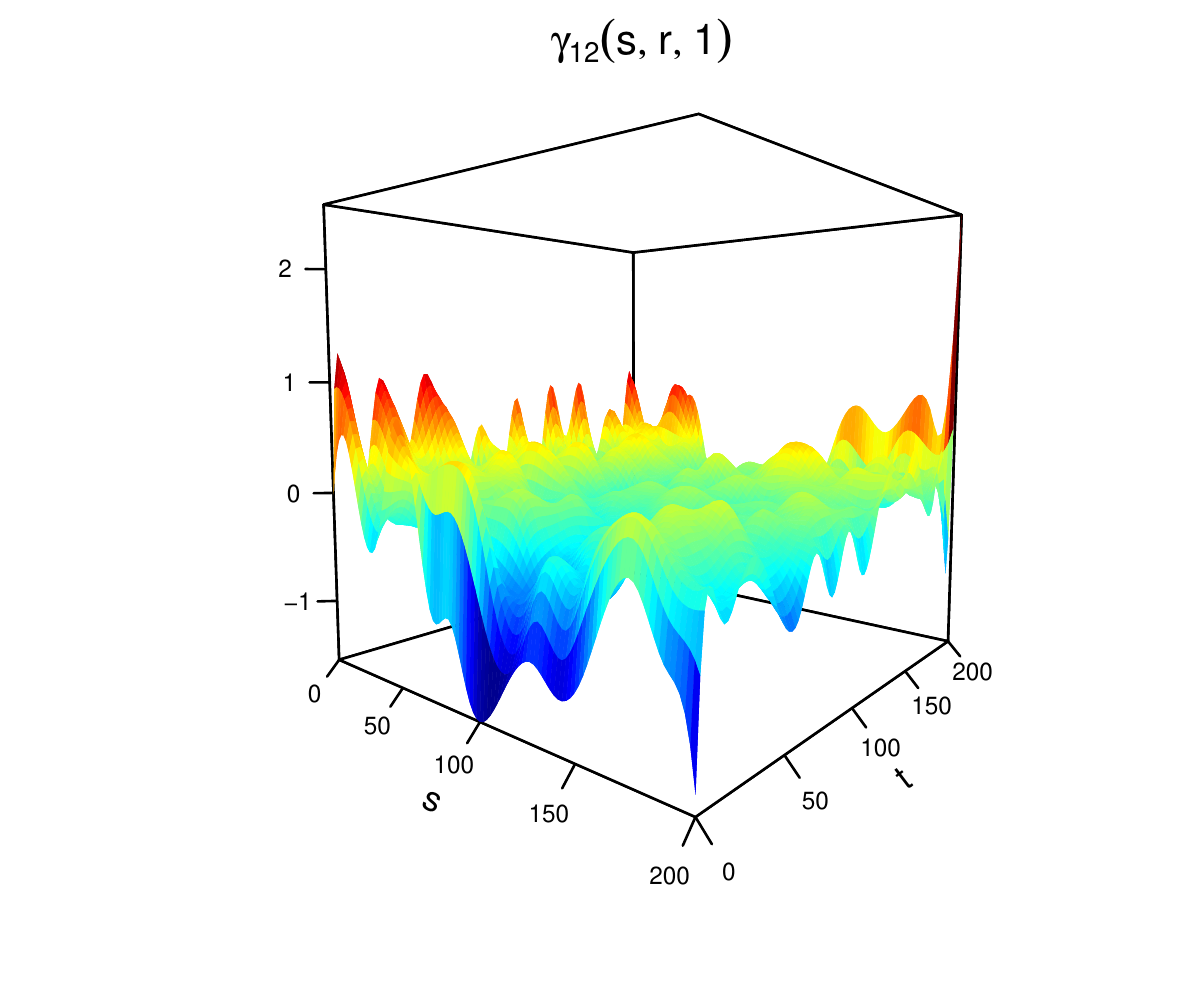}
  \caption{Estimated coefficient functions for the linear ($\beta_i(s,t)$ for $i = 1,2$), quadratic ($\gamma_{11}(s,r,t)$, at $t = 0$, only), and interaction ($\gamma_{12}(s,r,t)$, at $t = 0$, only) terms using the proposed PLS method for the Hawaii ocean data.}
  \label{fig:Fig_4}
\end{figure}

We show the estimated coefficient functions of the linear $\beta_1(s,t)$ and $\beta_2(s,t)$ and quadratic/interaction ($\gamma_{11}(s,r,t)$ and $\gamma_{12}(s,r,t)$ when $t = 0$, only) terms by our proposed method in Figure~\ref{fig:Fig_4}. The plots of $\beta_1(s,t)$ and $\beta_2(s,t)$ show that both potential density and temperature have greater impacts on salinity when $s < 25$ or $s > 175$. From the plots of $\gamma_{11}(s,r,t)$ and $\gamma_{12}(s,r,t)$, both potential density and temperature have negative quadratic/interaction effects on salinity at $t = 0$. 

\subsection{North Dakota weather data}

\begin{figure}[!htbp]
  \centering
  \includegraphics[width=5.7cm]{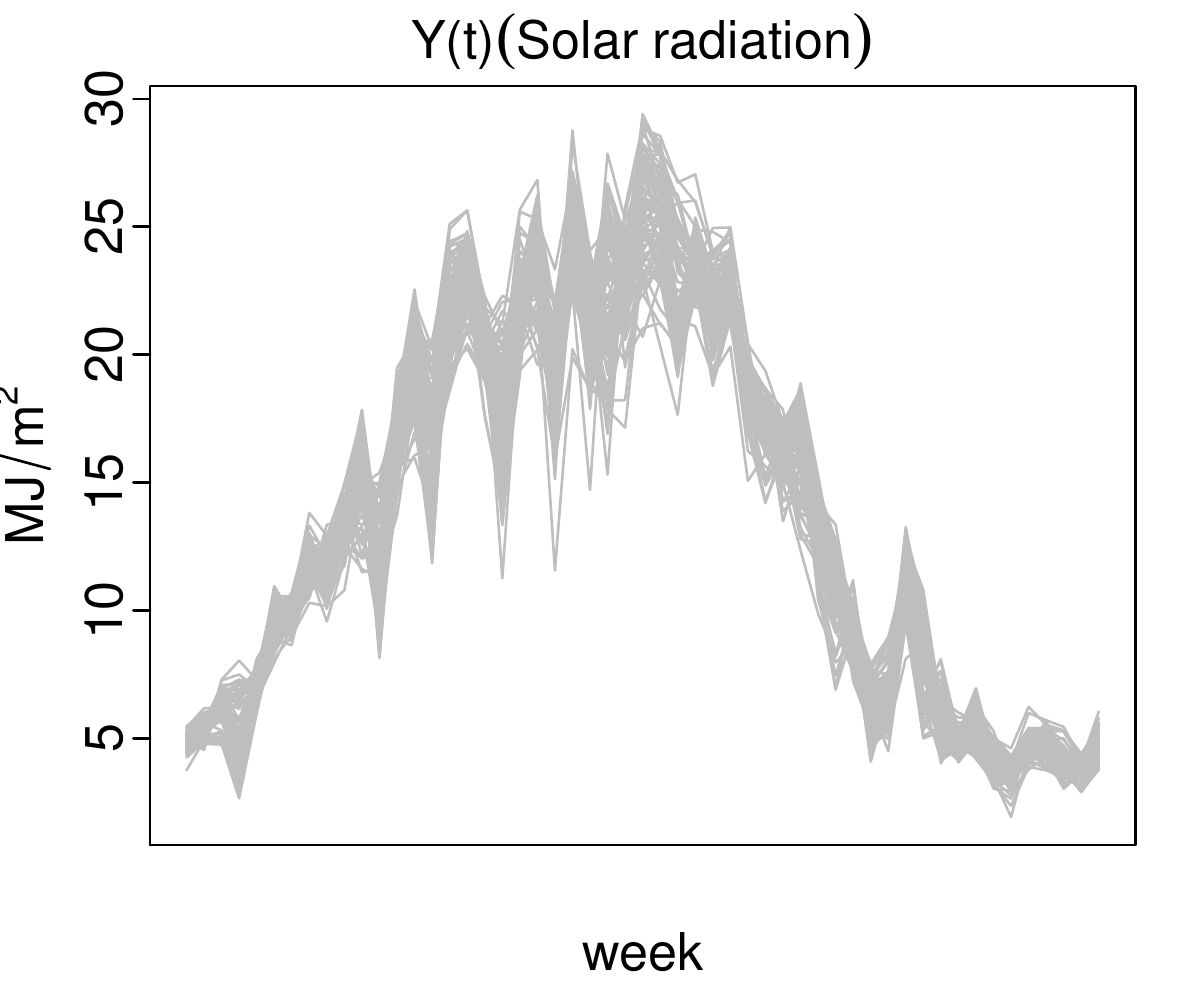}
  \includegraphics[width=5.7cm]{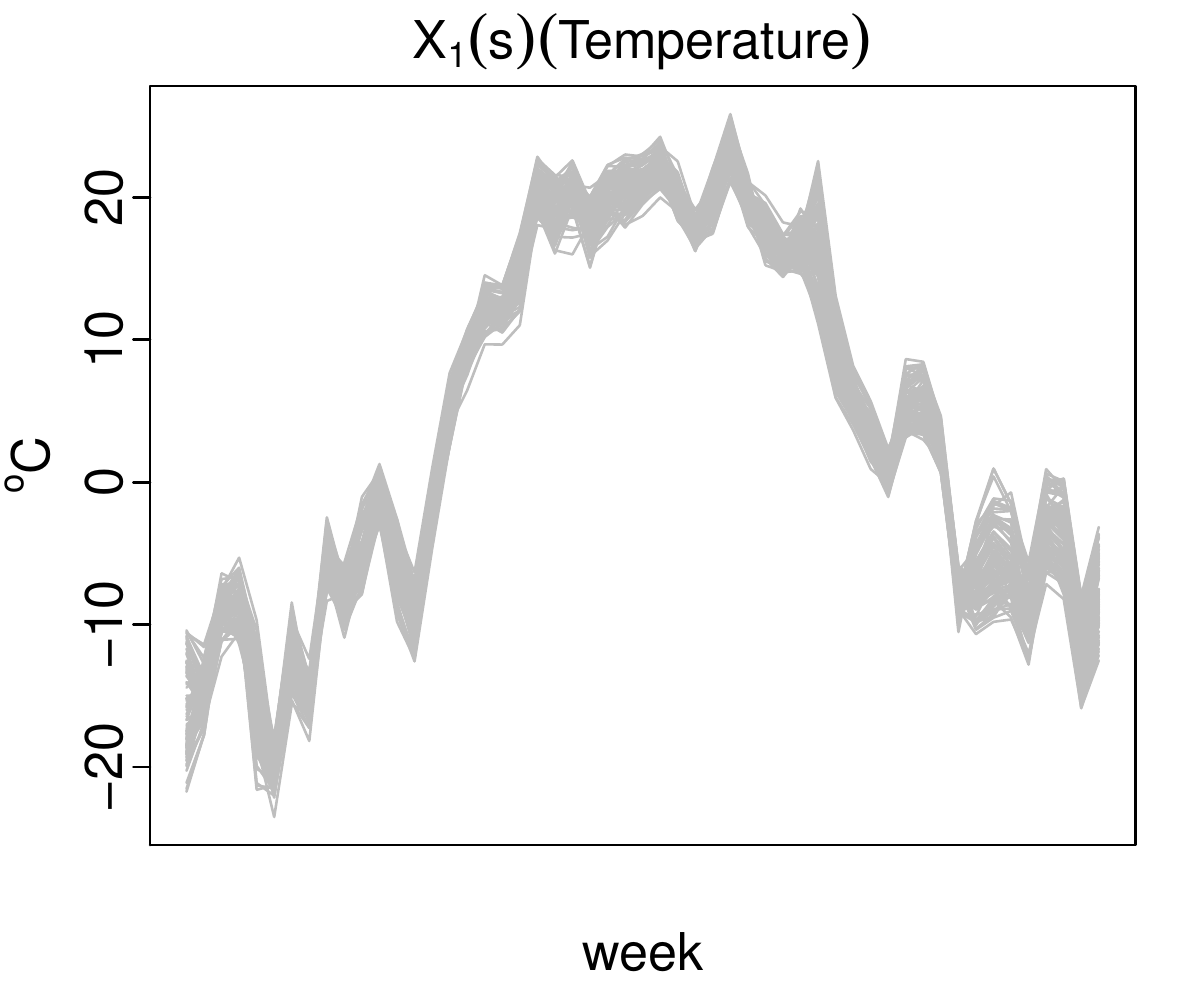}
  \includegraphics[width=5.7cm]{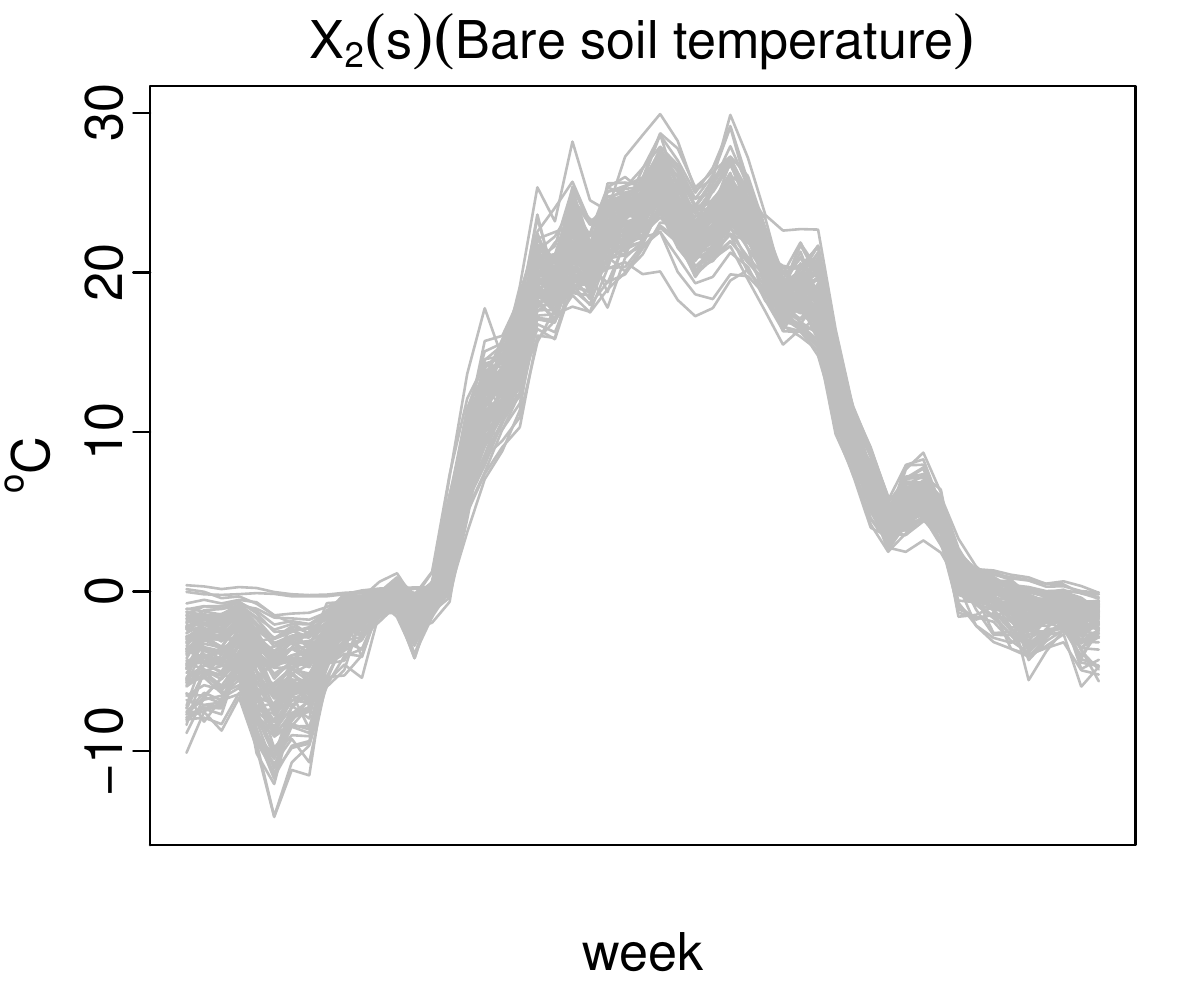}
\\  
  \includegraphics[width=5.7cm]{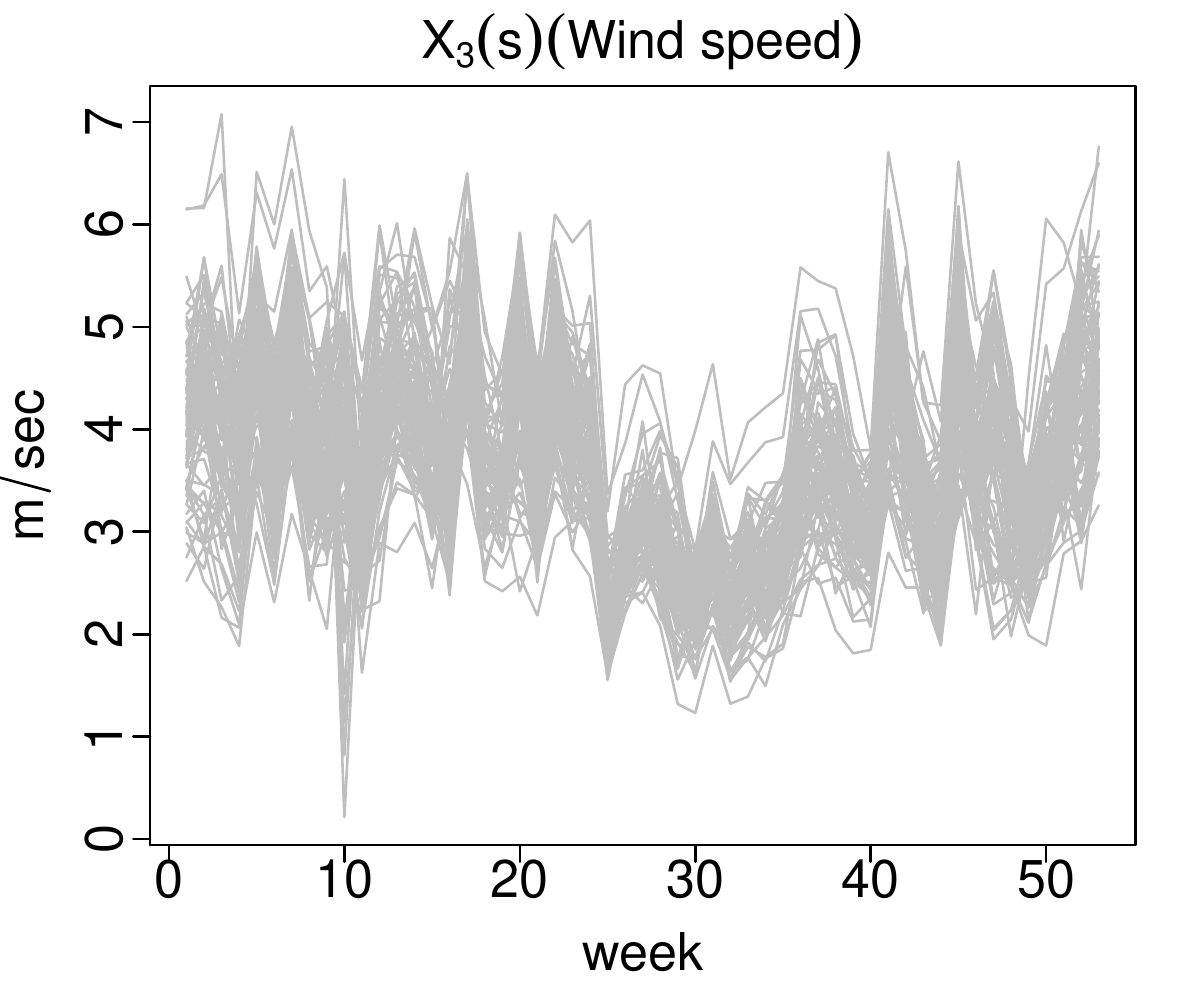}
    \includegraphics[width=5.7cm]{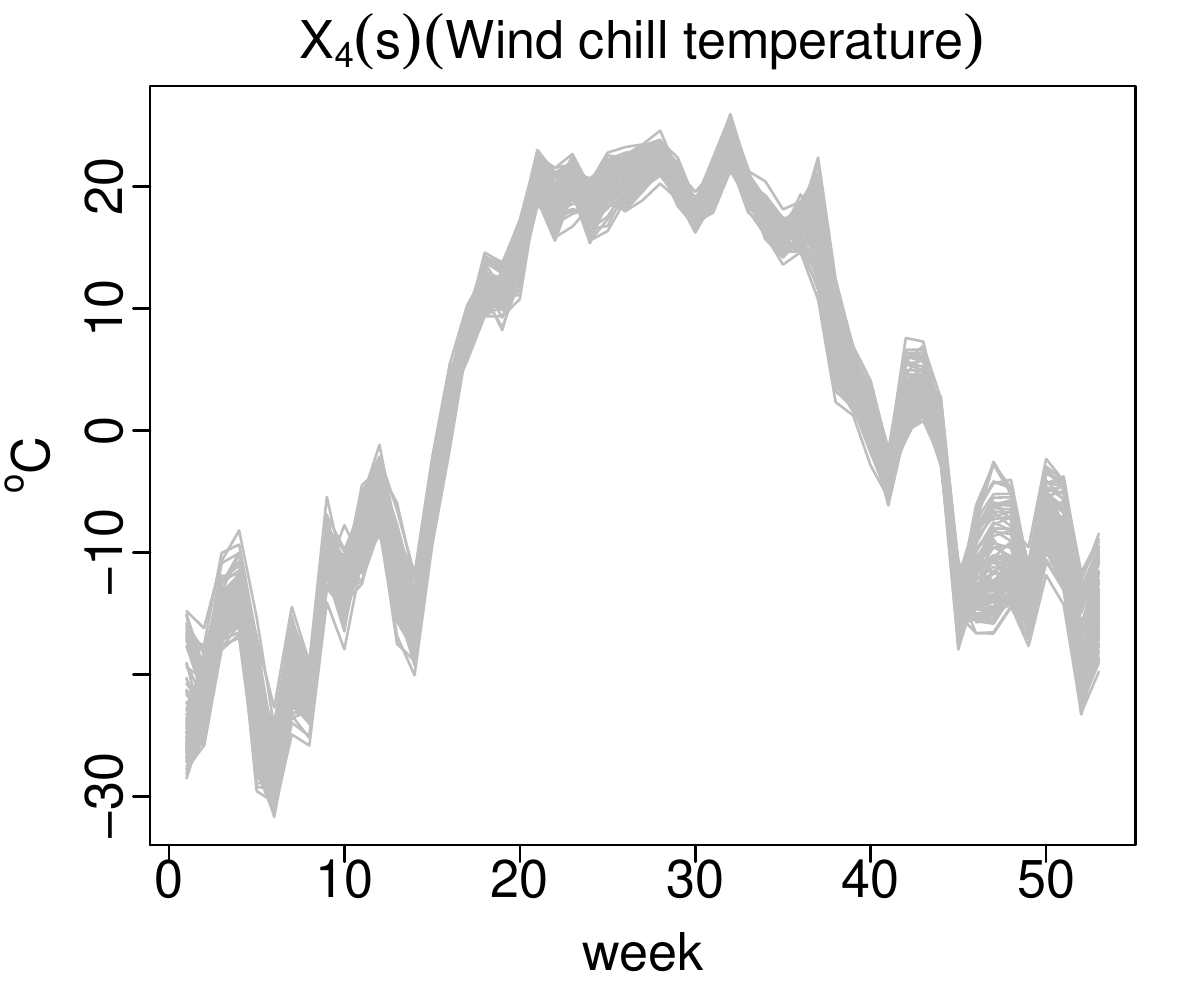}
  \caption{Plots of the functional variables; solar radiation, temperature, bare soil temperature, wind speed, and wind chill in the North Dakota weather dataset. The observations are the functions of weeks and $1 \leq s,t \leq 53$.}
  \label{fig:Fig_5}
\end{figure}

The North Dakota weather data (dataset is available from the North Dakota Agricultural Weather Network Center: \url{https://ndawn.ndsu.nodak.edu}) consists of five functional variables: solar radiation (MJ/m$^2$), temperature ($^\circ$C), bare soil temperature ($^\circ$C), wind speed (m/sec), and wind chill temperature ($^\circ$C). These variables are calculated by averaging daily values for the defined 7-day periods. The dataset were collected from 91 stations across North Dakota (see Table~\ref{tab:stations} in Appendix) from January~1, 2018, to December~31, 2018. For this dataset, the variables are viewed as functions of weeks, and there are 91 curves observed at 53 equally spaced points in the interval $[1, 53]$ for each variable. Herein, we considered predicting solar radiation using the other four variables. The plots of the functional variables are presented in Figure~\ref{fig:Fig_5}.

A similar procedure as that for the analysis of the Hawaii ocean data was followed to calculate the performance metrics and coefficients of determination values for both the LQ and proposed PLS methods. The dataset was randomly split into the following two parts: the model was constructed based on 50 randomly selected curves to predict remaining 41 curves for the solar radiation variable, and this process was repeated 100 times. For the proposed method, the performance metrics were calculated based on $K_{\Y} = K_{\pmb{\X}} = 40$ number of basis functions. The results are presented in Table~\ref{tab:pm_ND}. This table shows that the models obtained by the forward selection procedures of both the LQ and proposed methods outperform their full and main effect models. The coefficient of determination values produced by both methods are very close to 1, which indicates that both methods have good prediction performances. However, the results, reported in Table~\ref{tab:pm_ND}, show that the selected model of the proposed PLS method has better prediction performances compared with the selected model of LQ. 

\begin{table}[!htbp]
\tabcolsep 0.18in
\centering
\begin{small}
\caption{Computed average MSPE, RMSPE, MAPE, $R^2$, and $R^2_{pred}$ values of the LQ and PLS methods for the North Dakota weather data. The subscripts $_{\text{main}}$, $_{\text{full}}$, and $_{\text{selected}}$ respectively correspond to the case where the model is estimated using the main effect, full, and selected models. The values given in brackets are the estimated standard errors for the calculated performance metrics.}\label{tab:pm_ND}
\begin{tabular}{@{}llllrr@{}} 
\toprule
{Method} & {MSPE} & {RMSPE} & {MAPE} & {$R^2$} & {$R^2_{pred}$} \\
\midrule
LQ$_{\text{main}}$ 			& 1.445 (0.080)	& 0.094	(0.003)	& 0.069	(0.002)	& 0.988 (9.8 $\times 10^{-4}$)	& 0.986 (7.8 $\times 10^{-4}$) \\
LQ$_{\text{full}}$			& 1.422 (0.080)	& 0.093	(0.003)	& 0.068	(0.002)	& 0.989 (8.1 $\times 10^{-4}$)	& 0.986 (7.5 $\times 10^{-4}$) \\
LQ$_{\text{selected}}$		& 1.389 (0.103)	& 0.092	(0.004)	& 0.067	(0.002)	& 0.990 (8.6 $\times 10^{-4}$)	& 0.987 (9.7 $\times 10^{-4}$) \\
PLS$_{\text{main}}$ 		& 1.343 (0.078)	& 0.090	(0.003)	& 0.066	(0.002)	& 0.989 (1.1 $\times 10^{-3}$)	& 0.987 (7.6 $\times 10^{-4}$) \\
PLS$_{\text{full}}$			& 1.352 (0.078)	& 0.090	(0.003)	& 0.066	(0.002)	& 0.989 (1.1 $\times 10^{-3}$)	& 0.987 (7.8 $\times 10^{-4}$) \\
PLS$_{\text{selected}}$		& 1.300 (0.093)	& 0.087	(0.003)	& 0.064	(0.002)	& 0.990 (9.0 $\times 10^{-4}$)	& 0.989 (8.9 $\times 10^{-4}$) \\
\bottomrule
\end{tabular}
\end{small}
\end{table}

\begin{figure}[!htbp]
  \centering
  \includegraphics[width=8.8cm]{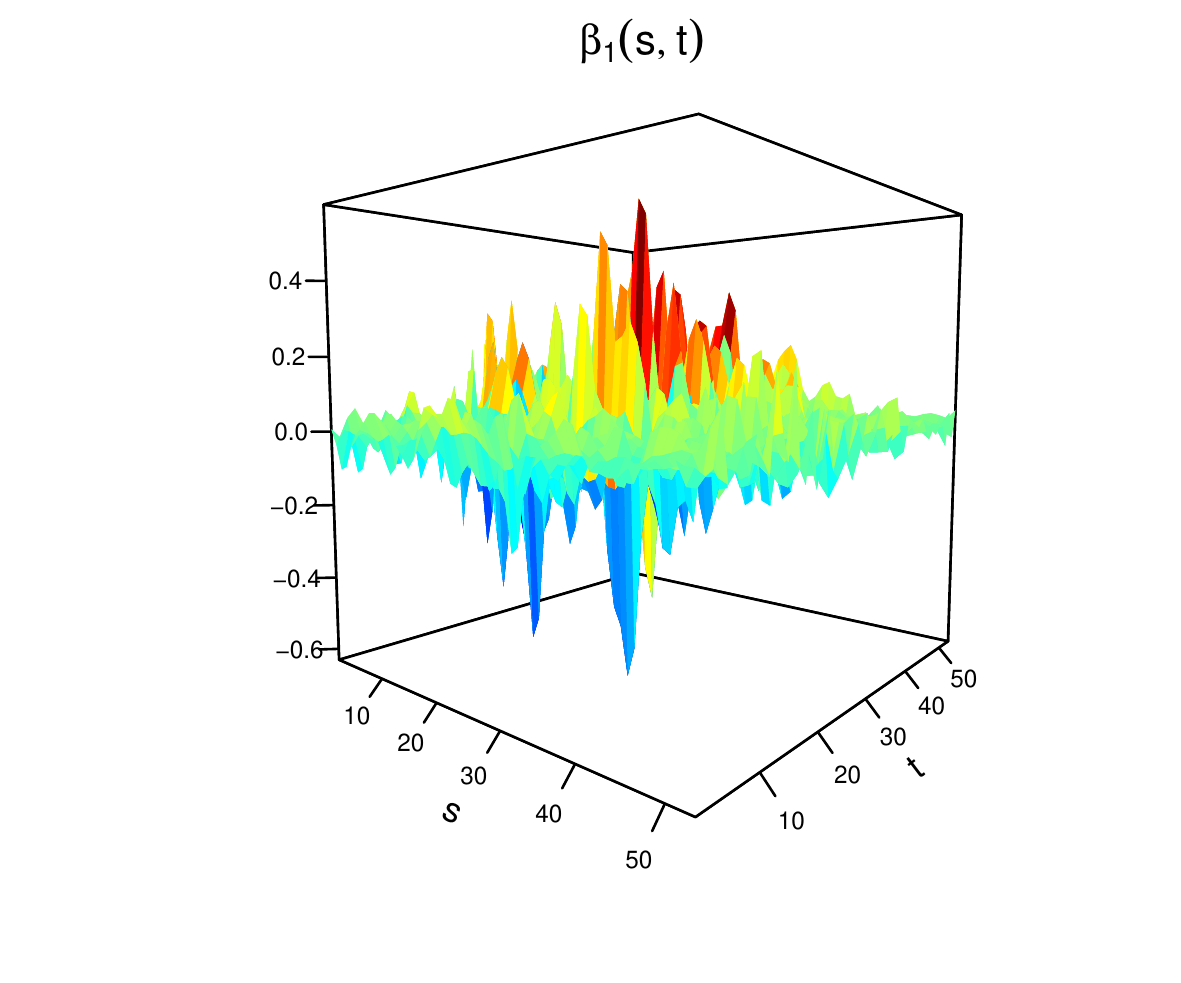}
\quad
  \includegraphics[width=8.8cm]{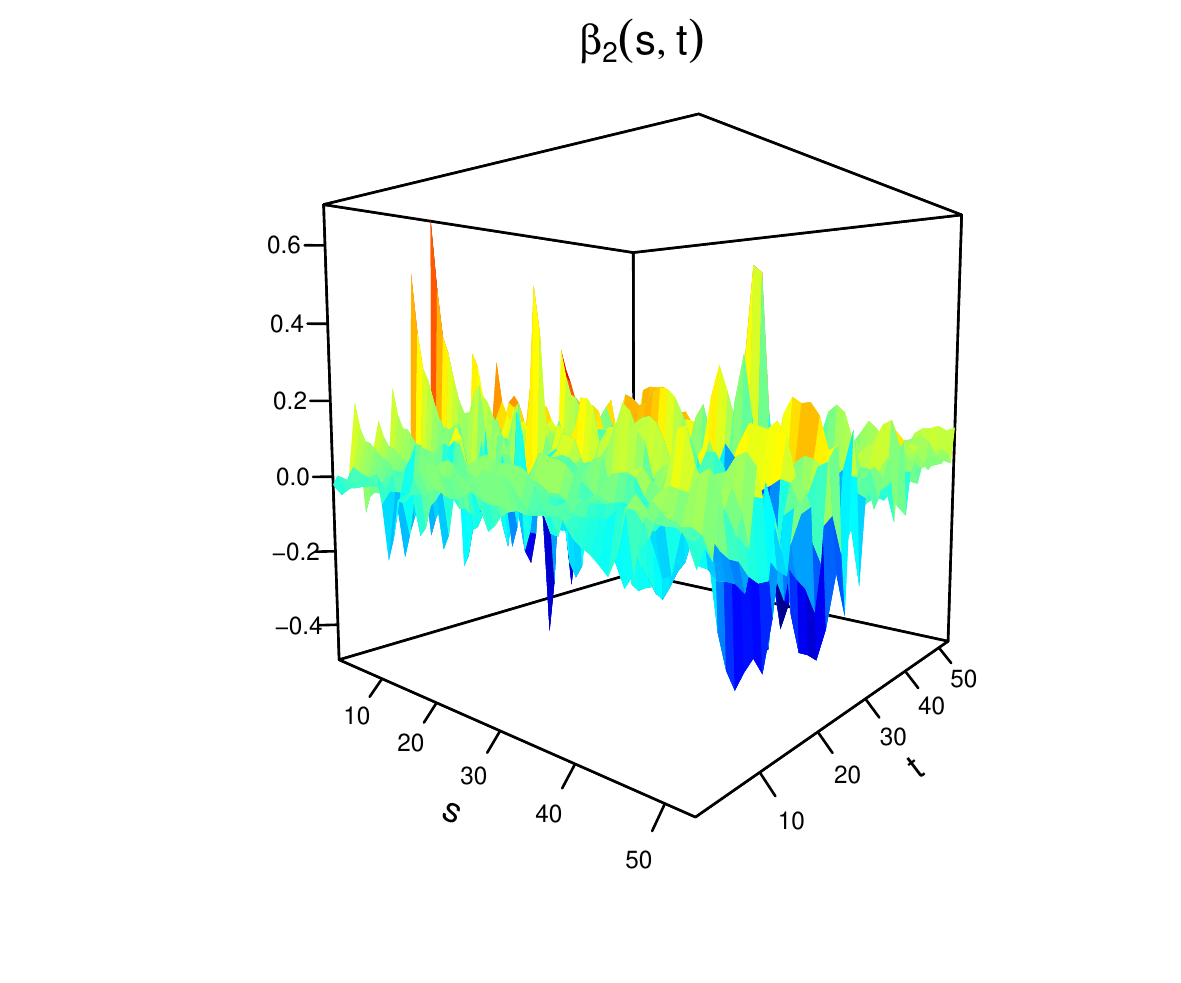}
\\  
  \includegraphics[width=8.8cm]{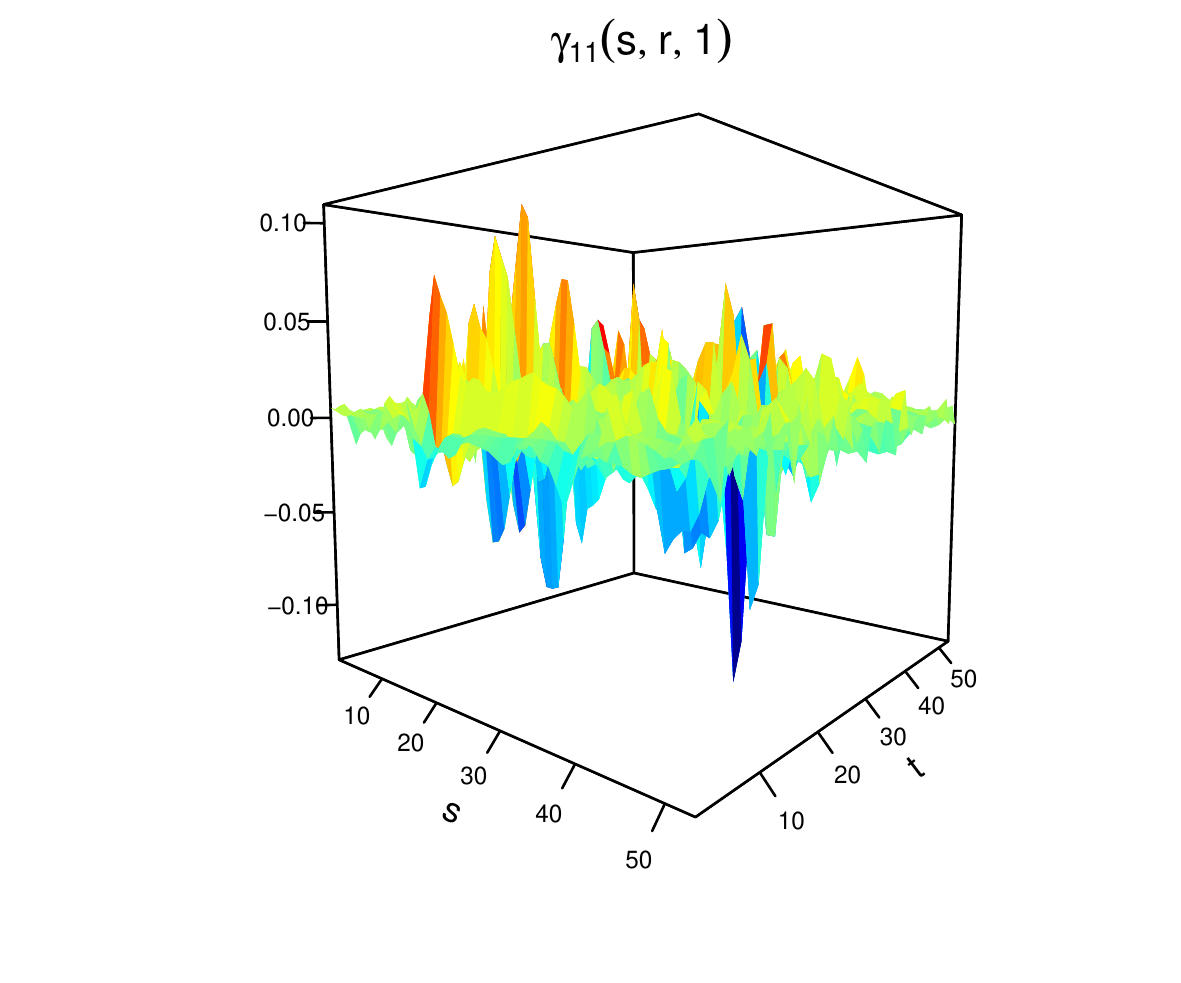}
\quad
    \includegraphics[width=8.8cm]{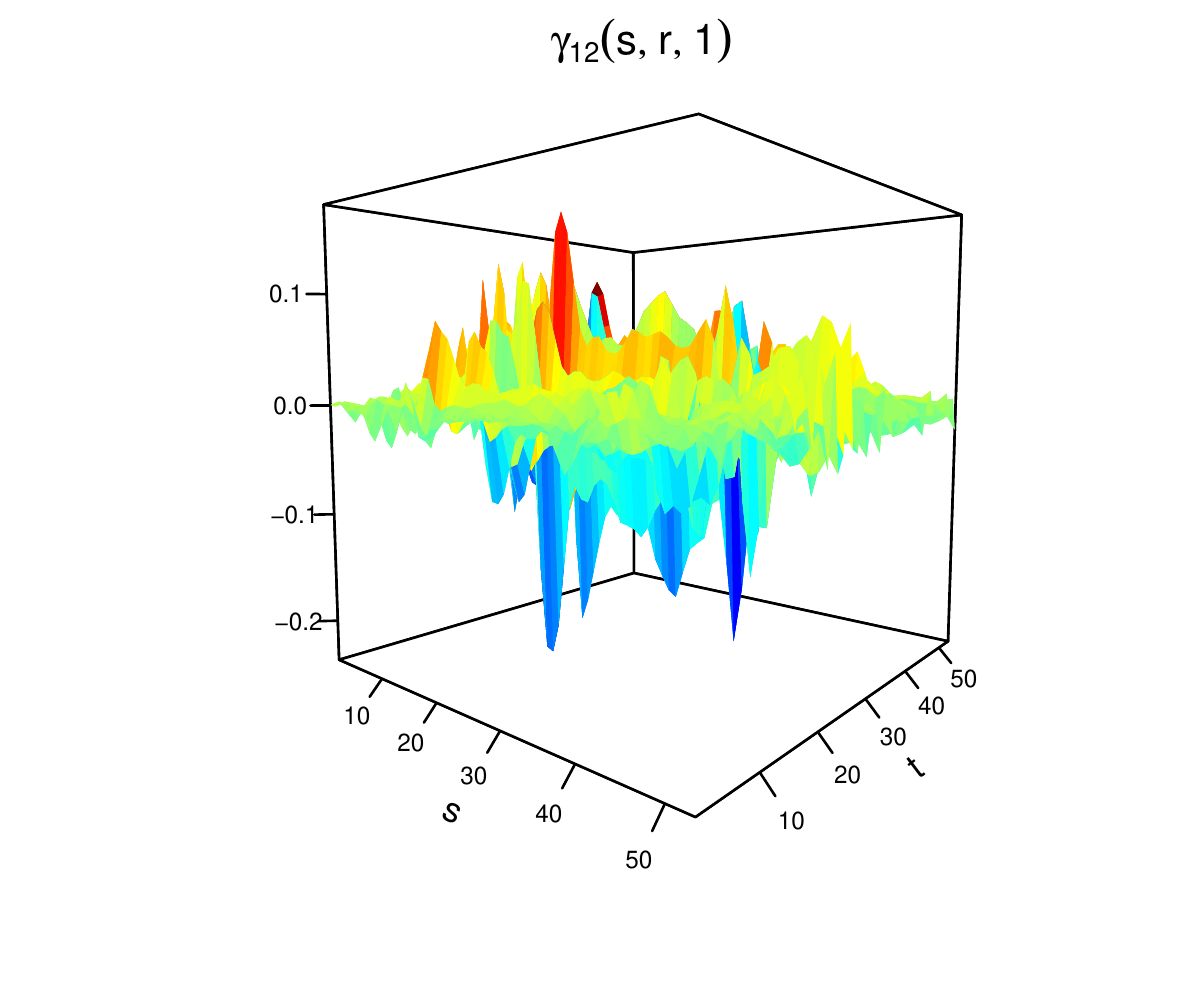}
  \caption{Estimated coefficient functions for the linear ($\beta_i(s,t)$ for $i = 1,2$), quadratic ($\gamma_{11}(s,r,t)$, at $t = 0$, only), and interaction ($\gamma_{12}(s,r,t)$, at $t = 0$, only) terms using the proposed PLS method for the North Dakota weather data.}
  \label{fig:Fig_6}
\end{figure}

For this dataset, a model using all 91 observations was constructed to determine the significant main and quadratic/interaction effect terms. Only the temperature $\X_1(s)$ and wind chill temperature $\X_4(s)$ were selected into the final model by both methods. The quadratic effect $\X_1(s) \X_1(r)$ was selected as significant by the LQ while the interaction effect $\X_1(s) \X_4(r)$ and the interaction effect $\X_4(s) \X_4(r)$ were selected as significant by the proposed PLS method. The estimated coefficient functions for the main $\beta(s,t)$ and quadratic/interaction ($\gamma(s,r,t)$ at $t = 1$, only) effects of the temperature and wind chill temperature are presented in Figure~\ref{fig:Fig_6}. It is clear from these plots that temperature and wind chill temperature have a greater effect on solar radiation when $s > 20$, and their quadratic and interactions have positive and negative quadratic effects along with $s$ and $r$ at $t = 0$.

\section{Conclusion} \label{sec:conc}

We propose a PLS method to estimate the function-on-function regression model where the multiple functional predictors have a quadratic term and interaction effects. In the proposed method, all the functional predictors and their quadratic/interaction effects are stacked into a vector, and the functional PLS regression of the functional response on the vectors of the functional predictors are considered. Since the direct estimation of a functional regression model is generally an ill-posed problem, the proposed method is estimated using the $B$-spline basis expansion of the functional objects. Basis expansion reduces the infinite-dimensional estimation problem to a simple finite-dimensional PLS regression setting using particular metrics in the spaces of the expansion coefficients. Using the metrics associated with the basis functions, the PLS regression constructed in this way is equivalent to the PLS regression constructed using a functional response and functional predictors. The finite sample performances of the proposed method are evaluated via several Monte Carlo experiments and two empirical data analyses. The performance of our method are compared with those of an existing method. Our results have shown that the proposed method produces improved prediction accuracy than those of the existing method. In addition, we propose a forward selection procedure to determine the significant main and quadratic/interaction effects. Our numerical results have proven that the forward selection procedure further improves the prediction performances of the PLS method when the true form of the model is even unspecified.

In the proposed method, we consider only $B$-spline basis expansion to approximate the functional objects. However, the specific forms of bivariate basis functions (probably tensor products of univariate B-splines) are generally unknown. In smoothing 2-dimensional surfaces, tensor-product type splines may ignore interior holes (or gaps) and yield extra bias as well as oversmoothing. Alternatively, the basis functions of non-tensor-product type for bivariate smoothing, such as Green's function \citep{Simonetal2008} and bivariate B-splines \citep{Huijun} may be used.

\newpage
\section*{Appendix}\label{sec:app}

\subsection*{Identifiability of the proposed model}\label{app_ident}
Model identifiability, which is an important theme in the estimation phase, for the function-on-function regression model with only main effects given in~\eqref{eq:fof} was theoretically discussed by \cite{Heetal2000} and \cite{Chiouetal2004}. Later, \cite{Scheipletal2016} discussed the identifiability of such models with realistic applications. As is stated in the Proposition 3.1 of \cite{Scheipletal2016}, the coefficient function $\beta_m(s,t)$ in~\eqref{eq:fof} is identifiable if and only if the kernel space of the covariance operator of the functional predictors ($K^{\X}$) is empty except the zero function, i.e., $ke(K^{\X}) = \lbrace 0 \rbrace$. \cite{LuoQi} extended the theoretical condition of model identifiability for function-on-function regression with only main effects to the model with interaction and quadratic effects. They defined the following $\lbrace p (p + 3)/2 \rbrace \times \lbrace p (p+3)/2 \rbrace$ dimensional covariance matrix of the predictor functions:
\begin{equation}\label{eq:cov}
\pmb{\Sigma} = 
\begin{bmatrix}
\pmb{A}(s,r) & \pmb{B}(s,r,u) \\
\pmb{B}^\top(s,r,u) & \pmb{C}(s, r, s^\prime, r^\prime)
\end{bmatrix},
\end{equation}
where the $p \times p$ and $p \times \lbrace p (p+1)/2 \rbrace$ dimensional submatrices $\pmb{A}(s,r)$ and $\pmb{B}(s,r,u)$ equal to $\text{Cov} \left( \X(s) \X(r) \right)$ and $\text{Cov} \left( \X(s), \X(r) \X(u) \right)$, respectively. The $\lbrace p (p+1)/2 \rbrace \times \lbrace p (p+1)/2 \rbrace$ dimensional matrix $\pmb{C}(s, r, s^\prime, r^\prime)$ equals to $\text{Cov} \left( \X(s) \X(r), \X(s^\prime) \X(r^\prime) \right)$. Then,  \cite{LuoQi} proved that the coefficients in model~\eqref{eq:main} are identifiable if the kernel space of $\pmb{\Sigma}$ in~\eqref{eq:cov} only has the zero function. Please see Proposition S.1. of \cite{LuoQi} for the proof. Based on the above, our proposed method given in~\eqref{main_ed} is also identifiable since the proposed method is a reformulated version of the model~\eqref{eq:main}.

\subsection*{Proof of Proposition~\eqref{prop:2}}\label{app_proof}
\begin{proof}%[Proof of Proposition~\eqref{prop:2}]
The proof of Proposition~\eqref{prop:2} is an extended version of the proof of Proposition 2 of \cite{Agu2010}. Denote by $\Lambda = \pmb{\Phi}^{1/2} \pmb{c}$ and $\Pi = \pmb{\Psi}^{1/2} \pmb{d}$ the column random vectors.

For any random variable $Z \in \mathcal{L}_2[0,1]$, the followings hold when $h = 1$:
\begin{align*}
W^{\Y} &= \int_0^1 \sum_{k=1}^{K_{\Y}} c_k \phi_k(t) \mathbb{E} \left[ \sum_{i=1}^{K_{\Y}} Z c_i \phi_i(t) \right] dt = \Lambda^\top \mathbb{E} \left[ \Lambda Z \right] = W^{\Lambda} Z, \\
W^{\pmb{\X}} &= \int_0^1 \int_0^1 \sum_{j=1}^{K_{\pmb{\X}}} \sum_{l=1}^{K_{\pmb{\X}}} d_{jl} \psi_{jl}(s,r) \mathbb{E} \left[ \sum_{\iota = 1}^{K_{\pmb{\X}}} \sum_{\tau = 1}^{K_{\pmb{\X}}} Z d_{\iota \tau} \psi_{\iota \tau}(s,r) \right] ds dr = \Pi^\top \mathbb{E} \left[ \Pi Z \right] = W^{\Pi} Z.
\end{align*}

The residuals obtained from the first iteration are given by:
\[ \begin{cases} 
      \Y_1(t) = \Y(t) - \zeta_1(t) \eta_1, & \zeta_1(t) \in \mathcal{L}_2[0,1] \\
      \Lambda_1 = \Lambda - \widetilde{\zeta}_1 \eta_1, & \widetilde{\zeta_1} \in \mathbb{R}^{K_{\Y}} 
   \end{cases}
\]

\[ \begin{cases} 
      \pmb{\X}_1(s,r) = \pmb{\X}(s,r) - p_1(s,r) \eta_1, & p_1(s,r) \in \mathcal{L}_2[0,1] \\
      \Pi_1 = \Pi - \widetilde{p}_1 \eta_1, & \widetilde{p}_1 \in \mathbb{R}^{K_{\pmb{\X}}^2} 
   \end{cases}
\]
Then, we have $\Lambda_1 = \pmb{\Phi}^{1/2} \pmb{c}_1$ and $\Pi_1 = \pmb{\Psi}^{1/2} \pmb{d}_1$, where $\pmb{c}_1$ and  $\pmb{d}_1$ are the random vectors of the basis coefficients of $\Y_1(t)$ and $\pmb{\X}_1(s,r)$, respectively. For functional PLS, the residuals calculated from the first iteration are as follows:
\begin{align*}
\Y_1(t) &= \sum_{k=1}^{K_{\Y}} c_{1,k} \phi_k(t) = \pmb{c}_1^\top \pmb{\Phi}(t), \qquad \forall t \in \mathcal{L}_2[0,1],\\
\pmb{\X}_1(s,r) &= \sum_{j=1}^{K_{\pmb{\X}}} \sum_{l=1}^{K_{\pmb{\X}}} d_{1,jl} \psi_{jl}(s,r) = \pmb{d}_1^\top \pmb{\Psi}(s,r), \qquad \forall s,r \in \mathcal{L}_2[0,1].	
\end{align*}
On the other hand, the residuals from the PLS regression are obtained as follows:
\begin{align*}
\Y_1(t) &= \pmb{\Phi}^\top(t) \pmb{c} - \eta_1 \frac{\mathbb{E} \left[ \pmb{\Phi}^\top(t) \pmb{c} \eta_1 \right] }{\mathbb{E} \left[ \eta_1^2\right]},\\
\pmb{\X}_1(s,r) &= \pmb{\Psi}^\top(s,r) \pmb{d} - \eta_1 \frac{\mathbb{E} \left[ \pmb{\Psi}^\top(s,r) \pmb{d} \eta_1 \right] }{\mathbb{E} \left[ \eta_1^2 \right]},
\end{align*}
\begin{align*}
\Lambda_1 &= \pmb{\Phi}^{1/2} \pmb{c} - \eta_1 \frac{\mathbb{E} \left[ \pmb{\Phi}^{1/2} \pmb{c} \eta_1 \right]}{\mathbb{E} \left[ \eta_1^2 \right] },\\
\Pi_1 &= \pmb{\Psi}^{1/2} \pmb{d} - \eta_1 \frac{\mathbb{E} \left[ \pmb{\Psi}^{1/2} \pmb{d} \eta_1 \right]}{\mathbb{E} \left[ \eta_1^2 \right]}
\end{align*}
and thus we have $\pmb{c}_1 = \pmb{c} - \eta_1 \frac{\mathbb{E} \left( \pmb{c} \eta_1 \right)}{\mathbb{E} \left[ \eta_1^2 \right]}$ and $\pmb{d}_1 = \pmb{d} - \eta_1 \frac{\mathbb{E} \left[ \pmb{d} \eta_1 \right]}{\mathbb{E} \left[ \eta_1^2 \right]}$, 
that is, $W^{\Y_1} = W^{\Lambda_1}$ and $W^{\pmb{\X}_1} = W^{\Pi_1}$. Now, assuming $W^{\Y_{\ell}} = W^{\Lambda_{\ell}}$ and $W^{\pmb{\X}_{\ell}} = W^{\Pi_{\ell}}$ for each $\ell \leq h$. Also, let us assume that $W^{\Y_{h+1}} = W^{\Lambda_{h+1}}$ and $W^{\pmb{\X}_{h+1}} = W^{\Pi_{h+1}}$. Then, at step $h+1$ the followings hold:
\[ \begin{cases} 
      \Y_{h+1}(t) = \Y(t) - \zeta_1(t) \eta_1 - \cdots - \zeta_h(t) \eta_h, & \zeta_i(t) \in \mathcal{L}_2[0,1] \\
      \Lambda_{h+1} = \Lambda - \widetilde{\zeta}_1 \eta_1 - \cdots - \widetilde{\zeta}_h \eta_h, & \widetilde{\zeta_i} \in \mathbb{R}^{K_{\Y}} 
   \end{cases}
\]

\[ \begin{cases} 
      \pmb{\X}_{h+1}(s,r) = \pmb{\X}(s,r) - p_1(s,r) \eta_1 - \cdots - p_h(s,r) \eta_h, & p_i(s,r) \in \mathcal{L}_2[0,1] \\
      \Pi_{h+1} = \Pi - \widetilde{p}_1 \eta_1 - \cdots - \widetilde{p}_h \eta_h, & \widetilde{p}_i \in \mathbb{R}^{K_{\pmb{\X}}^2} 
   \end{cases}
\]
As for $h=1$ and using the orthogonality of $\eta_i$ for $i = 1, \cdots, h$, it is shown that $\Lambda_{h+1} = \pmb{\Phi}^{1/2} \pmb{c}_{h+1}$ and $\Pi_{h+1} = \pmb{\Psi}^{1/2} \pmb{d}_{h+1}$, which concludes the proof.
\end{proof}

\subsection*{Station names for the North Dakota weather data}

\begin{table}[htbp]
\centering
\tabcolsep 0.11in
\caption{Station names for the North Dakota weather data.}
\begin{tabular}{@{}lllllll@{}}
\toprule
Station & Station & Station & Station & Station & Station & Station  \\
\midrule
Ada			& Carson 		& Eldred		& Harvey		& Linton	& Pekin 		& St. Thomas	\\
Alamo		& Cavalier		& Fargo			& Hazen			& Lisbon	& Perley		& Stephen		\\
Baker		& Cooperstown	& Fingal		& Hettinger		& Mandan	& Pillsbury		& Streeter		\\			
Beach		& Crary			& Finley		& Hillsboro		& Marion	& Plaza			& Tappen		\\	
Berthold	& Crosby 		& Forest River	& Hofflund		& Mavie		& Prosper		& Turtle Lake	 \\
Bottineau	& Dagmar		& Fort Yates	& Hope			& Mayville	& Redstone		& Ulen			 \\
Bowbells	& Dazey			& Fox	 		& Humboldt		& McHenry	& Robinson		& Wahpeton		 \\
Bowman		& Dickinson 	& Froid			& Inkster		& Michigan	& Rolla			& Warren		 \\	
Brampton	& Dooley		& Galesburg		& Jamestown		& Minot		& Roseau		& Watford City	 \\
Brorson		& Dunn		 	& Garrison		& Karlsruhe		& Mohall	& Ross			& Waukon		\\	
Campbell	& Edgeley		& Grafton		& Kennedy		& Mooreton	& Rugby			& Williams		\\
Cando		& Edmore		& Grand Forks	& Langdon		& Mott		& Sabin			& Williston		 \\	
Carrington 	& Ekre			& Greenbush		& Leonard		& Oakes		& Sidney		& Wishek		 \\	
\bottomrule
\end{tabular}
\label{tab:stations}
\end{table}

\clearpage
\bibliographystyle{agsm}
\bibliography{Interaction}

\end{document}